\documentclass[12pt]{article}

\usepackage[height=8.85in,width=6.45in]{geometry}

\usepackage[utf8]{inputenc}
\usepackage[T1]{fontenc}
\usepackage{amsmath}
\usepackage{amssymb}
\usepackage{mathtools}
\numberwithin{equation}{section}
\allowdisplaybreaks

\usepackage{slashed}
\usepackage{braket}
\usepackage[usenames,dvipsnames,svgnames,table]{xcolor}
\usepackage[colorlinks,citecolor=DarkGreen,linkcolor=FireBrick,urlcolor=FireBrick,linktocpage,breaklinks=true]{hyperref}
\usepackage{cite}
\usepackage{graphicx}
\usepackage{tikz}
\usepackage{tikz-cd}
\usepackage{times}
\usepackage[scaled]{couriers}

\usepackage{bm}
\usepackage{subfig}

\usepackage{xcolor}
\usepackage{mdframed}
\newenvironment{claim}{  \begin{mdframed}[linecolor=black!0,backgroundcolor=black!10]\noindent\itshape\ignorespaces}{\end{mdframed}}

\renewenvironment{figure}[1][]{
  \begin{originalfigure}[#1]
    \begin{mdframed}[linecolor=black!0,backgroundcolor=black!1]
}{
    \end{mdframed}
  \end{originalfigure}
}

\def\Nequals#1{$\mathcal{N}{=}#1$}
\def\bC{\mathbb{C}}
\def\bD{\mathbb{D}}
\def\bR{\mathbb{R}}
\def\bZ{\mathbb{Z}}
\def\cO{\mathcal{O}}
\def\Av#1{\mathop\mathrm{Av}(#1)}
\def\CP{\mathbb{CP}}
\def\IP{\mathrm{IP}}
\def\deg{\mathop\mathrm{deg}}
\def\diag{\mathop\mathrm{diag}}
\def\tr{\mathop\mathrm{tr}}
\def\vev#1{\langle#1\rangle}
\def\doubleunderline#1{\underline{\underline{#1}}}
\def\ii{\mathrm{i}}

\newcommand{\be}{\begin{equation}}
\newcommand{\ee}{\end{equation}}
\newcommand{\bea}{\begin{eqnarray}}
\newcommand{\eea}{\end{eqnarray}}

\begin{document}

\begin{titlepage}

\begin{flushright}
IPMU-19-0099
\end{flushright}

\vskip 4cm

\begin{center}

{\LARGE  Reflection groups and 3d $\mathcal{N}\ge $ 6  SCFTs}

\bigskip
\bigskip

Yuji Tachikawa and Gabi Zafrir

\bigskip

\begin{tabular}{ll}
 & Kavli Institute for the Physics and Mathematics of the Universe (WPI), \\
& University of Tokyo, Kashiwa, Chiba 277-8583, Japan
\end{tabular}

\vskip 1cm

\end{center}

We point out that the moduli spaces of all known 3d \Nequals8 and \Nequals6 SCFTs,
after suitable gaugings of finite symmetry groups,
have the form $\bC^{4r}/\Gamma$ where $\Gamma$ is a real or complex reflection group 
depending on whether the theory is \Nequals8 or \Nequals6, respectively.

Real reflection groups are either dihedral groups, Weyl groups, or two sporadic cases  $H_{3,4}$.
Since the BLG  theories and the maximally supersymmetric Yang-Mills theories correspond to dihedral and Weyl groups, 
it is strongly suggested that there are two yet-to-be-discovered 3d \Nequals8 theories for $H_{3,4}$.

We also show that all known \Nequals6 theories  correspond to complex reflection groups collectively known as $G(k,x,N)$.
Along the way, we demonstrate that two ABJM theories $(SU(N)_k\times SU(N)_{-k})/\bZ_N$ and $(U(N)_k\times U(N)_{-k})/\bZ_k$ are actually equivalent.

\end{titlepage}

\setcounter{tocdepth}{2}
\tableofcontents

\section{Introduction and summary}

\subsection{Brief summary}
Our aim in this paper is to demonstrate that 3d \Nequals8 and \Nequals6 superconformal field theories (SCFTs) can be usefully labeled by real and complex reflection groups, respectively.
In 3d, known \Nequals8 theories are either the low-energy limit of an \Nequals8 super Yang-Mills, or a Bagger-Lambert-Gustavsson theory \cite{Bagger:2007jr,Gustavsson:2007vu}.\footnote{%
Some \Nequals6 Lagrangian theories are known to enhance to \Nequals8 quantum mechanically. 
We will discuss them at length below, in Sec.~\ref{sec:longerintro} and in Sec.~\ref{sec:summary}.
}
Their moduli spaces (after suitable finite gaugings) have the form $\bC^{4N}/\Gamma$,
where $\Gamma$ is a Weyl group for the former,
and a dihedral group for the latter.
Both are examples of real reflection groups, which are either i) a Weyl group, ii) a dihedral group, or iii) the symmetry $H_3$ of the icosahedron  in $\bR^3$ or the symmetry $H_4$ of the 120-cell  in $\bR^4$.

Known \Nequals6 theories, in contrast, consist essentially of  the  $U(N)_k\times U(N)_{-k}$ Aharony-Bergman-Jafferis-Maldacena (ABJM) theories  \cite{Aharony:2008ug},
the $U(N+x)_k\times U(N)_{-k}$ Aharony-Bergman-Jafferis (ABJ) theories \cite{Aharony:2008gk}, the $SU(N)_k\times SU(N)_{-k}$ theories
and the $USp(2N)_k\times O(2)_{-2k}$ theories.
We will see below that, again after suitable finite gaugings,
their moduli spaces are of the form $\bC^{4N}/\Gamma$,
where $\Gamma$ is a complex reflection group known as $G(k,p,N)$, where $p$ is a divisor of $k$.

This suggests us first that there is a strong possibility that there are two yet-to-be-discovered 3d \Nequals8 theories associated to $H_3$ and $H_4$.
It also tells us that it would be worth while to look for 3d \Nequals6 theories associated to exceptional complex reflection groups other than those in the infinite series $G(k,p,N)$.

Below, we will first give a more detailed introductory narrative in Sec.~\ref{sec:longerintro},
and make precise how we assign the reflection group to a theory in Sec.~\ref{sec:rule-of-the-game}.
We then make in Sec.~\ref{sec:3d4d} some comments on the situation in 4d.
In Sec.~\ref{sec:org} we will describe how the rest of the paper is organized.

\subsection{A survey of known theories with 16 or 12 supercharges}
\label{sec:longerintro}
Multiple supersymmetry places various constraints on the structure of a quantum field theory\footnote{%
In this paper we only consider non-gravitational theories.}.
In particular, maximally supersymmetric theories, with 16 non-conformal supercharges, are so strongly constrained that we can now at least entertain the possibility of their classification in the future.
If we assume that there is a Lagrangian description manifesting all the supersymmetries, 
any maximally supersymmetric theory in dimension $\ge 4$ is the dimensional reduction of supersymmetric Yang-Mills theories in 10d for some gauge group $G$, first constructed in \cite{Brink:1976bc}.
When $d=4$ they give rise to the celebrated $d=4$ \Nequals4 super-Yang-Mills theories.
Another familiar fact is that in $d=4$, any Lagrangian \Nequals3 theory is so strongly constrained that it automatically has \Nequals4 supersymmetry.

We now know that there are a few highly supersymmetric theories which do not have any Lagrangian manifesting all supersymmetries.
For example, with  6d \Nequals{(2,0)} supersymmetry,
we do not have any interacting Lagrangian theory,
but the worldvolume theories on multiple M5-branes and other constructions provide concrete examples \cite{Witten:1995zh}.
It is now widely believed that they are labeled by a simply-laced Dynkin diagram, see e.g.~\cite{Henningson:2004dh}.
Another cases of interest are 4d \Nequals3 theories.
As already stated above, there are no genuinely \Nequals3 theories with a manifestly \Nequals3 Lagrangian,
but string theory constructions of nontrivial examples were found a few years ago in \cite{Garcia-Etxebarria:2015wns}.
This begs a natural question: is there any 4d \Nequals4 SCFT which is not an \Nequals4 super Yang-Mills? 

The situation in 3d looks much less settled. 
Let us first consider the maximally supersymmetric cases, i.e.~\Nequals8.
In contrast to $d>3$, known \Nequals8 theories can be put into two infinite series.
The first series consists of (the low energy limit of) \Nequals8 super Yang-Mills theories.
These are the natural continuation of the maximally supersymmetric theories in higher dimensions.
In addition to them, we have the second series, consisting of the Bagger-Lambert-Gustavsson (BLG) theories \cite{Bagger:2007jr,Gustavsson:2007vu}, 
which can be written \cite{VanRaamsdonk:2008ft} in the general structure of \Nequals3 superconformal Chern-Simons-matter systems \cite{Kao:1995gf,Kapustin:1999ha,Schwarz:2004yj,Gaiotto:2007qi} 
with gauge group $SU(2)_k\times SU(2)_{-k}$ and bifundamentals,
whose supersymmetry  enhances already at the Lagrangian level due to a cancellation.
We will give a more detailed review of known \Nequals8 theories in Sec.~\ref{sec:summary}.

The \Nequals7 theories are known to be automatically \Nequals8 even without a Lagrangian \cite{Bashkirov:2011fr,Bashkirov:2012rf}, so the next case to be discussed are \Nequals6 theories.
Here we meet the Aharony-Bergman-Jafferis-Maldacena (ABJM) theories \cite{Aharony:2008ug} and Aharony-Bergman-Jafferis (ABJ) theories \cite{Aharony:2008gk} which are the $U(N+x)_k\times U(N)_{-k}$ Chern-Simons-matter theories with $x=0$ and $x\neq 0$.
We then have the special unitary variant, $SU(N)_k\times SU(N)_{-k}$ Chern-Simons-matter theories.
The orthosymplectic variant, $USp(2N)_k\times O(M)_{-2k}$ Chern-Simons-matter theories, generically has only \Nequals5 but enhances to \Nequals6 when $M=2$.
They are known to exhaust the Lagrangian theories with manifest \Nequals6 supersymmetry,
up to a change in the abelian part of the gauge group \cite{Schnabl:2008wj}.

We note that the ABJM theories at $k=1,2$ and the ABJ theory with $(x,k)=(1,2)$ are known to enhance to \Nequals8 quantum mechanically,
and are believed to be equal to the low energy limit of the \Nequals8 super Yang-Mills with the gauge algebra of type $A_{N-1}$, $D_N$, $B_N=C_N$, respectively \cite{Kapustin:2010xq,Bashkirov:2010kz,Gang:2011xp}.
The BLG theories at $k=1,2,3,4$ are also believed to be equivalent to the infrared limit of \Nequals8 super Yang-Mills \cite{Lambert:2010ji,Bashkirov:2011pt,Agmon:2017lga} \footnote{%
For a nice summary, the readers are referred to a beautiful talk by C\'ordova at Strings 2018 \cite{CordovaTalk}.
}.
This confused situation of theories make us wonder: 
is there a principle which allows us to classify this zoo of theories? 
Have we essentially found all \Nequals8 theories?

\subsection{A classification scheme using reflection groups}
\label{sec:rule-of-the-game}

Here we would like to propose to use reflection groups as a useful label for these highly supersymmetric theories.
Supersymmetry guarantees that the moduli space of 3d \Nequals8 theories are of the form $\bR^{8r}/\Gamma$, where the action of $\Gamma$ is induced from a $\bR$-linear action of $\Gamma$ on $\bR^r$, so that it commutes with the $SO(8)$ R-symmetry.
Similarly, for 3d \Nequals6 theories, the moduli space is guaranteed to be of the form $\bC^{4r}/\Gamma$, where the action  of $\Gamma$ is induced from a $\bC$-linear action of $\Gamma$ on $\bC^r$, so that it commutes with the $SU(4)$ R-symmetry.
For a reason which we do not understand,
it turns out that, after suitable finite gaugings if necessary,
$\Gamma$ is always a reflection group\footnote{%
We provide the basics of the theory of reflection groups in Appendix.~\ref{sec:app}.
}, i.e.~a group generated by a reflection,
where a reflection on $\bC^r$ refers to a linear transformation which fixes a subspace $\bC^{r-1}$.
Concretely, $\Gamma$ is a Weyl group for \Nequals8 super Yang-Mills, and
a dihedral group for the BLG theory.
Both are real reflection groups.
For the ABJ(M) theory, the group is a complex reflection group $G(k,p,N)$,
where $p$ is a divisor of $k$.
The action of this group on $\bC^N$ parametrized by $(z_1,\ldots,z_N)$ is generated by
the symmetric group $S_N$ 
together with \begin{equation}
z_i \mapsto e^{2\pi\ii  p/k } z_i,\qquad \text{other $z_j$ fixed},
\end{equation} and \begin{equation}
(z_i,z_j) \mapsto (e^{2\pi\ii  /k} z_i, e^{-2\pi\ii /k} z_j), \qquad \text{other $z_\ell$ fixed}.
\end{equation} 

Before proceeding, we need to pause on the qualification we have repeatedly made that we need to perform suitable finite gaugings if necessary.
To see the necessity, one simply needs to consider 3d \Nequals8 super Yang-Mills with gauge group $SU(N)\rtimes \bZ_2$ where $\bZ_2$ acts by charge conjugation.
The group $\Gamma$ is then $S_N \times \bZ_2$, which is not a reflection group when $N>3$.
Another way to see the issue is to consider any 3d \Nequals6 theory.
Any such theory is known to have a flavor $U(1)$ symmetry \cite{Bashkirov:2011fr},
and acts on $\bC^{4r}/\Gamma$ by a scalar multiplication.
As there is no anomaly in a 3d $U(1)$ flavor symmetry\footnote{%
In 3d there is no anomaly associated to the anomaly polynomial,
but we need to worry about the global anomalies. 
In recent years a general theory of global anomalies was developed, e.g.~in \cite{Kapustin:2014dxa,Freed:2016rqq,Yonekura:2018ufj}.
According to this, the global anomaly of a $d$-dimensional theory with a global symmetry $G$ of a fermionic theory is characterized by the torsion part of the spin bordism group $\mathop{\mathrm{Tors}}\Omega^\text{spin}_{d+1}(BG)$.
This characterization includes not only the pure $G$ anomalies but also mixed $G$-gravitational anomalies.
For our present purpose we need to know the case $G=U(1)$, $d=3$, for which we can find $\mathop{\mathrm{Tors}}\Omega^\text{spin}_4(BU(1))=0$ e.g.~in \cite{Garcia-Etxebarria:2018ajm,Monnier:2018nfs}.
},
we can pick an arbitrary finite subgroup $\bZ_n$ of $U(1)$ and gauge it, 
without ruining the \Nequals6 supersymmetry.
The gauged theory then has $\bC^{4r}/\Gamma'$ where $\Gamma'=\Gamma\times \bZ_n$.
Again, for any complex reflection group $\Gamma$, $\Gamma\times \bZ_n$ with a large enough $n$ is not a complex reflection group.

To ameliorate the situation, we note the following.
Consider a theory $Q$ with a non-anomalous finite 0-form symmetry $G$. 
Then we can consider a new theory $Q'=Q/G$ obtained by gauging $G$.
$Q'$ is known to have a dual 1-form symmetry $\hat G$ \footnote{%
For the basics of the higher-form symmetries, see \cite{Kapustin:2014gua,Gaiotto:2014kfa}.
We note that $\hat G$ is a finite group when $G$ is abelian but is something more generalized when $G$ is non-abelian \cite{Bhardwaj:2017xup}.
It still holds that we can still gauge $\hat G$ to get the original theory back.},
so that $Q'/\hat G=Q/G/\hat G=Q$.
Let us say $Q$ is a parent of $Q'$ and $Q'$ is a child of $Q$.
By repeating this procedure, we have a large network of theories related to each other by a series of finite gaugings. 
Let us call all such theories relatives of $Q$.
We need to be careful that the relations are however not necessarily `linear', in the following sense.
A theory $Q$ can have two non-anomalous 0-form symmetries $G_1$ and $G_2$ but there can be mixed anomalies between them.
Then $Q$ can have two children $Q/G_1$ and $Q/G_2$.
Similarly a theory $Q'$ can have two non-anomalous 1-form symmetries $G_1$ and $G_2$ which have mixed anomalies.
Then $Q'$ can have two parents $Q'/G_1$ and $Q'/G_2$.
Therefore, there is no guarantee that there is a unique `oldest' ancestor or a unique `youngest' descendant among the relatives. 

To be more explicit, consider the case of a 3d gauge theory whose gauge Lie algebra is $\mathfrak{h}$.
To completely specify the gauge theory, we need to fix the Lie group $H$ whose Lie algebra is $\mathfrak{h}$.
This involves fixing the component $H_e$ connected to the identity, and then deciding which outer-automorphism of $H_e$ to gauge.
All this needs to be done in a way compatible with the matter content and the Chern-Simons level.
For example, consider the case when $\mathfrak{h}=\mathfrak{so}(2n)$,
with no Chern-Simons term.
In an $O(2n)$ gauge theory, the parity outer-automorphism of $SO(2n)$ is gauged.
We can `ungauge' it by gauging the dual 1-form symmetry,
resulting in an $SO(2n)$ theory, which is a parent of the $O(2n)$ theory.
When there is no matter field which transforms nontrivially under $-1\in SO(2n)$,
the gauge group can be chosen to be $SO(2n)/\mathbb{Z}_2$.
This theory is obtained from the $SO(2n)$ theory by gauging the $\mathbb{Z}_2$ 1-form symmetry \cite{Kapustin:2014gua}, and therefore is a parent of the $SO(2n)$ theory.
Therefore the $SO(2n)/\mathbb{Z}_2$ theory is the oldest ancestor among the theories discussed here.

Now we can phrase our observation in a precise manner:
\begin{claim}
For any \Nequals8 or \Nequals6 theory $Q$, one can pick a relative of $Q$ which is `locally oldest' (in the sense that it has no non-anomalous 1-form symmetry which can be gauged), so that its moduli space is given by $\bC^{4r}/\Gamma$ where $\Gamma$ is a real or complex reflection group, depending on the number of supersymmetries.
\end{claim}
We call $\Gamma$ a reflection group of $Q$. We shall sometime refer to a `locally oldest' relative simply as oldest for brevity sake, though it should be understood with the subtleties explained above.  

Note that at this level of generality, we have not eliminated the possibility that $Q$ can have more than one `locally oldest' relative whose reflection groups are different.
Therefore we cannot speak of \emph{the} reflection group of $Q$ yet. 

For \Nequals8, however, the inspection of the list of known \Nequals4 theories and various data computed for them reveal the following: 
\begin{claim}
For an \Nequals8 theory $Q$, there is always a unique oldest relative, so that we can refer to \emph{the} real reflection group $\Gamma$ associated to $Q$.
Furthermore, two \Nequals8 theories are relatives if and only if the associated reflection groups are the same.
\end{claim}

Therefore, the real reflection groups seem to provide a periodic table of \Nequals8 theories. It should be noted that this comes about from various dualities that conjecturally lead to many cases associated with the same reflection group being equal to one another. We shall summarize the current situation in that regard later in Sec.~\ref{sec:summary}.
 
As we mentioned, real reflection groups are one of the following: \begin{itemize}
\item A dihedral group $I_2(m)$, for which we have the BLG theory,
where $I_2(m)=\mathbb{Z}_{m} \rtimes \mathbb{Z}_2$ is the dihedral group of $2m$ elements.
\item A Weyl group $\mathcal{W}_G$, for which we have the low-energy limit of super Yang-Mills theory with gauge group $G$.
\item The symmetry $H_3$ of the icosahedron  in $\bR^3$ or the symmetry $H_4$ of the 120-cell  in $\bR^4$.
\end{itemize}
This strongly suggests us the following: \begin{claim}
There are two yet-to-be-discovered \Nequals8 theories whose reflection groups are $H_3$ and $H_4$.
\end{claim}
The authors have currently no idea how one might construct them, or one might disprove of their existence.

For the \Nequals6 theories, the situation does not seem to be as clear-cut.
For all known \Nequals6 theories, including the standard ABJM and ABJ theories,
we find at least one relative whose moduli space is of the form $\bC^{4r}/\Gamma$ 
by a complex reflection group $\Gamma$.
However, it is difficult to ascertain if this is the unique locally oldest ancestor,
because of the complicated multiple abelian factors a generic \Nequals6 theory can have.

We also find that the reflection group $\Gamma$ cannot distinguish the \Nequals6 theories as in \Nequals8 cases, from the following easy observation.
On one hand, as we will see below, $\Gamma$ for the ABJ theories $U(N+x)_k\times U(N)_k$,
for which it is known that $|x|\le N/2$,
is necessarily of the form $G(k,p,N)$ where $p$ is a divisor of $k$.
On the other hand, it is clear that the theories with the same $N$ and $k$ but with a different $x$ are never relatives, since the study of the leading correction to the $S^3$ free energy in the large $N$ limit using AdS/CFT \cite{Bergman:2009zh,Aharony:2009fc} shows that these theories have different $S^3$ free energies at order $N^{1/2}$, while a finite gauging cannot change that part of the $S^3$ free energy.
From the pigeonhole principle it then follows that there are some $x\neq x'$ which correspond to the same $p$.

Still, as we will see, the way we find the relative whose moduli space is of the form $\bC^{4r}/\Gamma$ with the complex reflection groups $\Gamma$ uses the \Nequals6 enhancement condition on the Chern-Simons levels in an essential way,
suggesting the close relationship between the \Nequals6 supersymmetry and the complex reflection groups.
It seems worthwhile to look for any putative \Nequals6 theory for which the associated complex reflection group $\Gamma$ is not one of the infinite series $G(k,p,N)$,
but one of the exceptionals $G_4$ to $G_{37}$.
Again, the authors do not have any idea how one might construct them, or one might disprove of their existence.

\subsection{Some comments on the situation in 4d}
\label{sec:3d4d}
Let us compare the situations in 3d and in 4d.
In 4d, all known \Nequals4 theories are super Yang-Mills theories for a gauge group $G$.
We can define the concept of the relatives as in 3d.
The oldest ancestor corresponds to taking $G$ to be connected and of the adjoint type,
for which the moduli space is $\bR^{6r}/\Gamma$ where $\Gamma$ is the Weyl group.
We also know that the group $\Gamma$ distinguishes the known 4d \Nequals4 theories:
the crucial point here is that the Montonen-Olive duality identifies the cases $G=SO(N+1)$ and $G=USp(2N)$ whose Weyl groups are identical.
The Weyl groups among the real reflection groups can be characterized by the condition that they are crystallographic, i.e.~that they preserve a lattice $\bZ^r \subset \bR^r$.

The moduli spaces of \Nequals3 theories of \cite{Garcia-Etxebarria:2015wns} were studied by \cite{Aharony:2016kai} and were shown to be of the form $\bC^{3r}/\Gamma$ where $\Gamma=G(k,p,N)$ with $k=3,4,6$.
They are again characterized among the more general $G(k,p,N)$ groups by the condition that they are crystallographic, i.e.~that they preserve a lattice $\bZ^{2r} \subset \bC^r$.

The crystallographic condition, both in \Nequals4 and \Nequals3 cases, can be understood as follows.
We can regard these theories as special \Nequals2 theories.
Then the group $\Gamma$ gives the monodromy group of the Seiberg-Witten fibration,
and needs to act as a part of the electromagnetic duality group of the low-energy $U(1)^r$ theory.
Therefore it needs to preserve the electromagnetic charge lattice, leading to the said condition \cite{Caorsi:2018zsq}.

The observations so far suggests the following schematic picture:
\begin{equation}
\begin{array}{c|ccc}
& \mathcal{N}=4 & & \mathcal{N}=3 \\
\hline
\text{4d}&
\begin{array}{c}
\text{real crystallographic}\\
\text{reflection (=Weyl) groups}\\
\text{(found: 1977\cite{Brink:1976bc})}\end{array}
 &\subset& 
\begin{array}{c}
\text{complex crystallographic}\\
\text{reflection groups}\\
\text{(found: 2015\cite{Garcia-Etxebarria:2015wns})}\end{array}\\
&\cap &&\cap \\
\text{3d}&
\begin{array}{c}
\text{real reflection groups}\\
\text{(found: 1977\cite{Brink:1976bc}, 2007\cite{Bagger:2007jr,Gustavsson:2007vu} )}
\end{array}
 & \subset & 
\begin{array}{c}
\text{complex reflection groups}\\
\text{(found: 2008\cite{Aharony:2008ug,Aharony:2008gk})}\\
\end{array}
 \\
\hline
& \mathcal{N}=8 & & \mathcal{N}=6
\end{array}
\end{equation}

Before leaving this section, we provide two remarks.
\begin{itemize}
\item In a recent paper \cite{Argyres:2019ngz} the constraints on the moduli space of rank-2 \Nequals3 theories were studied very carefully.
There, it was found that there can be cases where the moduli space is of the form $\bC^{3\cdot 2}/\tilde\Gamma$ which is \emph{not} a discrete quotient of $\bC^{3\cdot 2}/\Gamma$ for any complex reflection group $\Gamma$.

If a rank-2 theory whose moduli space is $\bC^{3\cdot 2}/\tilde\Gamma$ actually exists, 
then such a theory cannot be associated to any complex reflection group.
This will disprove the 4d \Nequals3 version of our conjecture.
We can then compactify the 4d theory on $S^1$ and flow to the infrared limit.
Most probably, this will provide a 3d \Nequals6 theory whose moduli space is $\bC^{4\cdot2}/\tilde \Gamma$ which cannot be labeled by a complex reflection group, disproving the 3d \Nequals6 version of our conjecture.

That said, it is not at all clear that such a 4d theory actually exists.
The authors of the present paper prefer to be agnostic, and would like to take the position that our observation makes this question simply more interesting.
\item A chiral algebra, or equivalently a vertex operator algebra, can be associated to any 4d \Nequals2 SCFT \cite{Beem:2013sza}.
When the 4d supersymmetry is \Nequals3 or \Nequals4, the chiral algebra has \Nequals2 or small \Nequals4 super-Virasoro subalgebra. 
In \cite{Bonetti:2018fqz} \Nequals2 and small \Nequals4 chiral algebras were constructed for arbitrary complex reflection groups and real reflection groups, respectively,
without the crystallographic condition.
Their construction  reproduced known chiral algebras constructed in \cite{Beem:2013sza} for \Nequals4 super Yang-Mills and in \cite{Nishinaka:2016hbw} for the \Nequals3 theories of \cite{Garcia-Etxebarria:2015wns}.
Their result is in a sense too good, since for 4d theories we definitely need the crystallographic groups.
Hopefully, their chiral algebras for non-crystallographic real and complex reflection groups are somehow related to 3d \Nequals8 and \Nequals6 theories, which can conjecturally be usefully labeled by the same reflection groups.
\end{itemize}

\subsection{Organization of the rest of the paper}
\label{sec:org}
The rest of the paper is organized as follows. In Sec.~\ref{sec:summary} we summarize the known 3d \Nequals8 SCFTs and their relation with real reflection groups.
In Sec.~\ref{sec:moduli},
we study the moduli space of known \Nequals8 and \Nequals6 theories in detail.
We not only study the standard BLG, ABJM and ABJ theories,
but also study the most general versions where the gauge group contains multiple abelian factors.

Along the way, we will see that 
two versions of ABJM theories, $(U(N)_k \times U(N)_{-k}) /\bZ_k$  and $(SU(N)_k\times SU(N)_{-k})/\bZ_{N}$, have the same moduli space $\bC^{4N}/\Gamma$ where $\Gamma=G(k,k,N)$.
In Sec.~\ref{sec:equivalence},
we will show that these two theories are actually equivalent, by slightly extending the argument of \cite{Witten:2003ya}.
We also explicitly check the agreement of their superconformal indices.

We note that in \cite{Lambert:2010ji} the agreement of the moduli spaces was established
and that in \cite{Honda:2012ik} the agreement of the superconformal indices was checked only when $k$ and $N$ are coprime.
This was due to their assumption that the $\bZ_k$ part acts diagonally as a subgroup of the $U(1)$ baryonic symmetry,
without mixing with the gauge group.
Our $\bZ_k$ action is more general and therefore our result does not contradict theirs.
We also note that our results here overlaps with \cite{Bergman:20XX}.

Finally, we provide the basics of reflection groups in Appendix~\ref{sec:app}.
We do not claim any originality in the appendix; we simply hope that the contents might be of some use to the readers.

\section{Summary of known 3d \Nequals8 oldest SCFTs}
\label{sec:summary}

\subsection{The table}
We shall first begin by summarizing the known interacting 3d \Nequals8 SCFTs, particularly concentrating on the oldest members, and the relationships between them. As mentioned in the introduction, the known 3d \Nequals8 SCFTs are either the low-energy limit of super Yang-Mills (SYM) theories, the BLG theories or the ABJ(M) type $U(N)_1 \times U(N)_{-1}$, $U(N)_2 \times U(N)_{-2}$ and $U(N+1)_2 \times U(N)_{-2}$ theories. Out of this, for the super Yang-Mills theories the cases where the group is connected and of adjoint-type are oldest, while for the BLG theories the oldest are the ones with gauge group $(SU(2)_k\times SU(2)_{-k})/\bZ_{2}$. As will be discussed in the next section, for the ABJ(M) type theories the oldest are $U(N)_1 \times U(N)_{-1}$, $(U(N)_2 \times U(N)_{-2})/\bZ_{2}$ and $U(N+1)_2 \times U(N)_{-2}$. All these have the moduli space $\bC^{4r}/\Gamma$ for $\Gamma$ a real reflection group, where several cases have the same moduli space. However, it is known that many of these cases are dual to each other so the actual list of distinct oldest SCFTs is smaller. In fact the dualities are such that the known theories are consistent with the distinct \Nequals8 oldest SCFTs being labeled by real reflection groups. The purpose of this section is to summarize these relations, and point out the additional dualities necessary for this conjecture to hold. We have summarized this information in table \ref{tableSum}. 

\begin{table}
\vbox{}
\vspace*{-1cm}
\[
\begin{array}{|c|c|c|c|}
\hline
\text{real reflection}&\text{SYM type}&\text{BLG type}& \text{ABJ(M) type}\\
\text{group}&\text{SCFTs}&\text{SCFTs}& \text{SCFTs}\\
\hline
\hline
\mathcal{W}_{SU(2)} = \bZ_2 & \frac{SU(2)}{\bZ_2}& \IP[\frac{SU(2)_1\times SU(2)_{-1}}{\bZ_2} ] & \IP[ U(2)_1\times U(2)_{-1} ],\\
 & &  & U(2)_2\times U(1)_{-2}\\
\hline
\mathcal{W}_{Spin(4)} = \bZ_2 \times \bZ_2 & \frac{Spin(4)}{\bZ_2 \times \bZ_2}& \frac{SU(2)_2\times SU(2)_{-2}}{\bZ_2} & \frac{U(2)_2\times U(2)_{-2}}{\bZ_2}\\
\hline
\mathcal{W}_{SU(3)} = S_3 & \frac{SU(3)}{\bZ_3}& \frac{SU(2)_3\times SU(2)_{-3}}{\bZ_2} & \IP[U(3)_1\times U(3)_{-1}],\\
&&& \frac{U(2)_3\times U(2)_{-3}}{\bZ_3}\\
\hline
\mathcal{W}_{USp(4)} = I_2(4) & \frac{USp(4)}{\bZ_2}& \frac{SU(2)_4\times SU(2)_{-4}}{\bZ_2} & U(3)_2\times U(2)_{-2},\\
&&& \frac{U(2)_4\times U(2)_{-4}}{\bZ_4}\\
\hline
\mathcal{W}_{SU(4)} = S_4 & \frac{SU(4)}{\bZ_4}& & \IP[U(4)_1\times U(4)_{-1}],\\
 &  & & \frac{U(3)_2\times U(3)_{-2}}{\bZ_2}\\
\hline
\mathcal{W}_{SU(N)} = S_N, & \frac{SU(N)}{\bZ_N}& & \IP[U(N)_1\times U(N)_{-1}]\\
N>4 & & & \\
\hline
\mathcal{W}_{Spin(2N)} = G(2,2,N), & \frac{Spin(2N)}{Z_{Spin(2N)}}& & \frac{U(N)_2\times U(N)_{-2}}{\bZ_2}\\
N>3 & & & \\
\hline
\mathcal{W}_{USp(2N)} = \mathcal{W}_{Spin(2N+1)} = & \frac{USp(2N)}{\bZ_2}, & & U(N+1)_2\times U(N)_{-2}\\
G(2,1,N), N>2 & \frac{Spin(2N+1)}{\bZ_2}& & \\
\hline
\mathcal{W}_{G_2} = I_2(6) & G_2& \frac{SU(2)_6\times SU(2)_{-6}}{\bZ_2} & \frac{U(2)_6\times U(2)_{-6}}{\bZ_6}\\
\hline
\mathcal{W}_{F_4}  & F_4& & \\
\hline
\mathcal{W}_{E_6}  & \frac{E_6}{\bZ_3}& & \\
\hline
\mathcal{W}_{E_7}  & \frac{E_7}{\bZ_2}& & \\
\hline
\mathcal{W}_{E_8}  & E_8& & \\
\hline
I_2(m), m\neq 2,3,4,6 & & \frac{SU(2)_m\times SU(2)_{-m}}{\bZ_2} &  \frac{U(2)_m\times U(2)_{-m}}{\bZ_m}\\
\hline
H_3  & & & \\
\hline
H_4  & & & \\
\hline
\end{array}
\]
\caption{List of the real reflection groups and the \Nequals8 SCFTs realizing them broken into three categories: SYM type, BLG type or ABJ(M) type. Here only oldest SCFTs are listed. We use $\mathcal{W}_{G}$ for the Weyl group of $G$, and $Z_{Spin(2N)}$ for the center of $Spin(2N)$, which is either $\bZ_2 \times \bZ_2$ or $\bZ_4$ depending on whether $N$ is even or odd. Also we use $\IP[x]$ for the interacting part of the SCFT $x$. $I_2(m)$, $H_3$ and $H_4$ are the non-crystallographic real reflection groups, as explained in the introduction. An empty entry implies no representative SCFT in this category. As noted in the text, there is by now some evidence that all theories in a given line in fact describe the same SCFT. There are no known \Nequals8 SCFTs associated with the real reflection groups $H_3$ and $H_4$, but we have kept them in the table for completeness.
\label{tableSum}}
\end{table}

\subsection{Comments}
We next review some of the entries in the table. 
\subsubsection{ABJM vs.~SYM of type A}
 First the ABJM model, $U(N)_1\times U(N)_{-1}$, describes the physics of $N$ M2-branes and so should flow to the same SCFT as the maximally supersymmetric $U(N)$ theory. This has been checked by matching the sphere partition function\cite{Kapustin:2010xq} and superconformal index\cite{Gang:2011xp}. For super Yang-Mills theories the calculation of these quantities is generally hindered by the fact that the resulting expression, evaluated from the gauge theory Lagrangian using the localization results, diverges. This is usually attributed to the full $SO(8)$ R-symmetry not being fully manifest in the UV Lagrangian. In the case of the $U(N)$ theory one can use a dual description, which is essentially the same theory but with the addition of a fundamental $U(N)$ hyper, to calculate these quantities.

The $U(N)_1\times U(N)_{-1}$ theory is known to contain a decoupled part, associated with the $U(1)_1\times U(1)_{-1}$ theory, which is just a free \Nequals8 SCFT (containing a free \Nequals4 hyper and twisted hyper or 8 massless real scalars and 8 massless Majorana fermions)\cite{Bashkirov:2010kz}. The reminder is an interacting SCFT which we shall denote by $\IP[U(N)_1\times U(N)_{-1}]$. On the $U(N)$ SYM side, this is mapped to the $U(1)$ part being decoupled, and flowing to the free \Nequals8 SCFT. Since $U(N)= \frac{U(1)\times SU(N)}{\bZ_N}$, the interacting part is expected to be that of $\frac{SU(N)}{\bZ_N}$ SYM. This then leads to the duality between the interacting part of the $U(N)_1\times U(N)_{-1}$ theory and the SCFT associated with $\frac{SU(N)}{\bZ_N}$ SYM\cite{Gang:2011xp}.

\subsubsection{ABJ(M) vs.~SYM of type BCD}

The other ABJM and ABJ theories which enhance to \Nequals8 are also expected to be dual to SYM theories of type $SO$ and $USp$. This is again motivated by string theory as these theories should describe M2-branes on an OM2 plane. Here checking partition functions is harder due to the aforementioned problem. However, several tests have been done in \cite{Gang:2011xp} and these suggest that the other ABJM theory, $U(N)_2\times U(N)_{-2}$, is dual to adjoint-type $O(2N)$ SYM and that the ABJ representative, $U(N+1)_2\times U(N)_{-2}$, is dual to $SO(2N+1)$ and $\frac{USp(2N)}{\bZ_2}$ \footnote{The moduli space of the SYM theories at the UV is really of the form $(\bR^7 \times S^1)^{r}/\Gamma$, and only becomes $\bC^{4r}/\Gamma$ at low energies. There is a subtlety though in this low-energy limit as points which were at finite distance before can be separated infinitely far apart. When we consider the SYM theories here we always consider the theories at the origin of the Coulomb branch, but it is possible for a flow from a different point to lead to a different SCFT. For instance it is argued in \cite{Gang:2011xp}, from string theory reasonings, that this happens for the $usp(2N)$ SYM theories, and that performing the flow from the antipodal point leads instead to the $U(N)_2\times U(N)_{-2}$ SCFT.}. 

Before mentioning the tests used for this proposal, we want to elaborate about the $U(N)_2\times U(N)_{-2}$ theory and the $O$ SYM theory. The SYM theories with gauge groups $SO(2N+1)$, $USp(2N)$ and $O(2N)$ have the same moduli space, which differs from that of gauge group $SO(2N)$ \footnote{The groups $SO(2N+1)$, $USp(2N)$ and $SO(2N)$ all generically have the automorphism group $G(2,1,N)$. However, while for $SO(2N+1)$ and $USp(2N)$ all the automorphisms are inner automorphisms, and therefore part of the Weyl group, for $SO(2N)$ a $\bZ_2$ subgroup is outer leading to a smaller Weyl group. The outer automorphism element acts on the $SO(2N)$ group by the exchange of its two spinor representations, which is the same way parity acts on it. As a result in the $O(2N)$ group this element becomes an inner automorphism.}. Notably the former have at low energies the moduli space $\bC^{4N}/G(2,1,N)$ while the latter has the moduli space $\bC^{4N}/G(2,2,N)$. The ABJM theory $U(N)_2\times U(N)_{-2}$ has the moduli space $\bC^{4N}/G(2,1,N)$ and so can be identical to the $O$ SYM theory and not the $SO$. However, has we shall show in the next section, the $U(N)_2\times U(N)_{-2}$ is a child of the `locally oldest' $\frac{U(N)_2\times U(N)_{-2}}{\bZ_2}$, whose moduli space is $\bC^{4N}/G(2,2,N)$. Since the two are related by gauging a discrete symmetry, it is natural to conjecture that the $\frac{U(N)_2\times U(N)_{-2}}{\bZ_2}$ SCFT is the same as adjoint-type $Spin(2N)$ SYM theory.

Back to the equivalence between the ABJ(M) and SYM theories of type BCD, the results were motivated by various tests preformed by \cite{Gang:2011xp}. These tests are more intricate then the ones for the $U(N)$ case, where they again rely on adding $m$ fundamental hypers though in this case there is no duality to the cases without the fundamental matter. However, the addition can be mapped to changing the M2-brane background by an additional $\bZ_m$ quotient, which they then mapped to the AdS dual side and used it, together with knowledge regarding the behavior of the added states to match indices at large $N$. They also used various low-rank coincidences to check superconformal indices at low $N$, and then rely on flows to connect this with the large $N$ analysis. This last part will be mostly of interest to us here. 

It is known that $SU(2)=USp(2)=Spin(3)$, $Spin(4)=SU(2)\times SU(2)$ and that $Spin(6)=SU(4)$. It was then noted in \cite{Gang:2011xp} that the index of the $U(2)_2\times U(1)_{-2}$ theory and the interacting part of $U(2)_1\times U(2)_{-1}$ match and that the indices of $U(2)_2\times U(2)_{-2}$ and $U(3)_2\times U(3)_{-2}$ match a $\bZ_2$ gauging of the square of the interacting part of $U(2)_1\times U(2)_{-1}$ and the interacting part of $U(4)_1\times U(4)_{-1}$, respectively. This matches with the expected low-rank coincidences using the duality between the interacting part of the ABJM theories and $\frac{SU(N)}{\bZ_N}$ SYM theories. The reason why we stress this is that it suggests how the duality should work at the group level rather than just the algebra level. The end result then is that the $\frac{U(N)_2\times U(N)_{-2}}{\bZ_2}$ SCFT is expected to be dual to the $Spin(2N)$ SYM of adjoint-type, while the $SO(2N+1)$ and $USp(2N)$ SYM theories of adjoint-type should be dual to one another and to the $U(N+1)_2\times U(N)_{-2}$ SCFT.      

\subsubsection{BLG vs.~SYM}
This brings us to the BLG theories, which have the moduli space $\bC^{8}/I_2(m)$, where $I_2(m)$ is the ordinary dihedral group of order $2m$. Here $m=2k$ for the $SU(2)_k\times SU(2)_{-k}$ and $m=k$ for the $(SU(2)_k\times SU(2)_{-k})/\bZ_{2}$ variant, which is the oldest one. For generic values of $k$ these give different moduli spaces than those of the ABJM, ABJ and SYM theories. However, for $k=1,2,3,4$ and $6$ there are some equivalences among the moduli spaces. For $k=1,2,3,4$ this is thought to be due to dualities, which we shall next review. 

The $(SU(2)_1\times SU(2)_{-1})/\bZ_{2}$ SCFT is thought to be dual to the ABJM $U(2)_1\times U(2)_{-1}$ theory and so should contain a decoupled free sector and an interacting part which should be equivalent to the $\frac{SU(2)}{\bZ_2}$ SYM theory. The $SU(2)_2\times SU(2)_{-2}$ SCFT is thought to be dual to the ABJM $U(2)_2\times U(2)_{-2}$ and so also to the $\frac{O(4)}{\bZ_2\times \bZ_2}$ SYM theory. As a result the $(SU(2)_2\times SU(2)_{-2})/\bZ_{2}$, $(U(2)_2\times U(2)_{-2})/\bZ_{2}$ and the $\frac{SO(4)}{\bZ_2\times \bZ_2}$ SYM theory should also be dual. The duality between the first two is just a special case of a more general duality which will be discussed in Sec.~\ref{sec:equivalence}. The $(SU(2)_4\times SU(2)_{-4})/\bZ_{2}$ SCFT is thought to be dual to the ABJ $U(3)_2\times U(2)_{-2}$ theory, and so also to the $\frac{USp(4)}{\bZ_2}$ SYM. These three dualities were first proposed in \cite{Bashkirov:2011pt}. Additionally, the $(SU(2)_3\times SU(2)_{-3})/\bZ_{2}$ SCFT is thought to be dual to the interacting part of the ABJM $U(3)_1\times U(3)_{-1}$ theory and so also to the $\frac{SU(3)}{\bZ_3}$ SYM theory. This duality was first proposed in \cite{Agmon:2017lga}. In all these cases the dualities between the BLG and ABJ(M) theories can be checked by the computation and matching of partition functions.

This leaves us with the $k=6$ case. In this case we have that $I_2(6)=\mathcal{W}_{G_2}$, and so BLG theories with this moduli space share it with the low-energy limit of $G_2$ SYM theory. There are two BLG theories with this moduli space, $(SU(2)_6\times SU(2)_{-6})/\bZ_{2}$ and $SU(2)_3\times SU(2)_{-3}$ out of which only the former is `locally oldest'. The latter theory is expected to be a $\bZ_{2}$ gauging of the $(SU(2)_3\times SU(2)_{-3})/\bZ_{2}$ SCFT, which is thought to be dual to the $\frac{SU(3)}{\bZ_3}$ SYM theory. Indeed, the SYM theory has a $\bZ_{2}$ discrete symmetry acting as charge conjugation on the $su(3)$ gauge algebra, and gauging it leads to the moduli space $\bC^{8}/I_2(6)$. 

This brings us to the $(SU(2)_6\times SU(2)_{-6})/\bZ_{2}$ SCFT. We can ask whether this SCFT and the low-energy limit of the $G_2$ SYM theory are the same theory. If 3d \Nequals8 SCFTs are in one to one correspondence with real reflection groups then this must hold. There is indeed some indirect evidence for this which we shall next present. This relies on an observation in \cite{Aharony:2016kai} regarding 4d \Nequals3 SCFTs. Specifically, following the construction of 4d \Nequals3 SCFTs using S-folds in \cite{Garcia-Etxebarria:2015wns}, \cite{Aharony:2016kai} studied some of their moduli spaces. The notable observation that will be of interest to us here is that some cases appear to have an enhancement of supersymmetry to \Nequals4. This is motivated by the appearance of a Coulomb branch operator of dimension two in these cases, which from superconformal representation theory must be accompanied with the additional supercurrents. 

This happens for three cases, and the resulting theories are consistent with being just \Nequals4 SYM theories with gauge algebras $su(3)$, $usp(4)$ and $g_2$. When compactified to 3d, this class of theories are known to give the ABJM theories. Specifically, the three cases with the \Nequals4 enhancement should reduce to the ABJM theories $(U(2)_k\times U(2)_{-k})/\bZ_{k}$ for $k=3,4$ and $6$. Here we have used the structure of the moduli space, which will be discussed in detail in the next section, to determine the exact group structure. Next we can use the duality between $(U(2)_k\times U(2)_{-k})/\bZ_{k}$ and $(SU(2)_k\times SU(2)_{-k})/\bZ_{2}$, which will be discussed in Sec.~\ref{sec:equivalence}, to map the resulting theories to the BLG cases instead. As we previously mentioned the $k=3,4$ and $6$ cases were found to be consistent with the \Nequals4 SYM theories with gauge algebras $su(3)$, $usp(4)$ and $g_2$, respectively. Assuming this is true, we are led to identify the $(SU(2)_k\times SU(2)_{-k})/\bZ_{2}$ BLG theory with the \Nequals8 SYM theory associated with $su(3)$ for $k=3$, $usp(4)$ for $k=4$ and $g_2$ for $k=6$, up to some choice of group structure. Finally, we note that the $k=3$ and $k=4$ cases just reproduce some of the dualities we discussed before. The $k=6$ case is new and suggests the equivalence of the $(SU(2)_6\times SU(2)_{-6})/\bZ_{2}$ SCFT and the low-energy limit of the $G_2$ SYM theory. 

As we mentioned previously, we will argue in Sec.~\ref{sec:equivalence} that the BLG $(SU(2)_k\times SU(2)_{-k})/\bZ_{2}$ theory and the ABJM type $(U(2)_k\times U(2)_{-k})/\bZ_{k}$ theory are dual. This provides a dual description for all the oldest BLG theories. We also note that for the $k=4$ case there should be an additional dual in the orthosymplectic family. It was suggested in \cite{CGHNP}, following an observation in \cite{Aharony:2008gk}, that the $O(2N)_2 \times USp(2N)_{-1}$ and $U(N)_4\times U(N)_{-4}$ theories are dual. Specifically for $N=2$, this suggests that the $O(4)_2 \times USp(4)_{-1}$ theory has a $\bZ_4$ 1-form symmetry, and gauging it should lead to the SCFT associated with the $(SU(2)_4\times SU(2)_{-4})/\bZ_{2}$ theory. Since this theory lies outside the known families of \Nequals8 SCFTs, we have not written it in the table.

\subsubsection{Exceptional theories}
Finally we have the remaining exceptional maximally supersymmetric gauge theories, $f_4$, $e_6$, $e_7$ and $e_8$. These realize the real reflection groups that are just the Weyl groups of these algebras. There are also the two exceptional real reflection groups $H_3$ and $H_4$ for which there is currently no known 3d \Nequals8 SCFTs. It is an interesting question whether \Nequals8 SCFTs realizing these moduli spaces exist or not. Another interesting question arising from this discussion is the calculation of partition function, like the $S^3$ or superconformal index, for maximally supersymmetric Yang-Mills theories. Besides the obvious use in checking many of the dualities summarized here, we think it is also of physical interest to calculate this for all \Nequals8 SCFTs, like the ones associated with exceptional groups.   

\section{Moduli spaces of known theories}
\label{sec:moduli}
\subsection{\Nequals8 super Yang-Mills theory}
Let us first recall the moduli space of the \Nequals8 super Yang-Mills theory whose gauge group is $G$.
Let us assume $G$ is connected. 
We denote its rank by $r$.

The vector multiplet contains seven scalar fields $\phi^{I=1,\ldots,7}$ in the adjoint $\mathfrak{g}$ of $G$.
On a generic point of the moduli space, $\phi^I$ all commute, and take values in the Cartan subalgebra $\mathfrak{h}:=\bR^r \subset \mathfrak{g}$.
This breaks $G$ to the Cartan subgroup $T:=U(1)^r$.
Abelian gauge fields in 3d can be dualized to periodic scalars, which are parameterized by the torus $\hat T$ dual to $T$.
Finally we need to take into account the action of the Weyl group $\Gamma$, resulting in the moduli space of the form \begin{equation}
(\mathfrak{h}^7 \times \hat T)/\Gamma
\end{equation} of dimension $8r$.
Note that this moduli space depends on $G$ not just on $\mathfrak{g}$.

To take the low energy limit, one needs to pick a point on this moduli space. Choosing the origin, the moduli space of the low energy SCFT is given by \begin{equation}
\bR^{8r} / \Gamma.
\end{equation} 
We note that it is independent of the choice of the connected group $G$ belonging to the same algebra $\mathfrak{g}$.

We can also consider possibly disconnected group $\tilde G$ containing outer automorphisms of $\mathfrak{g}$ as the gauge group.
In such cases the discrete identification $\Gamma$ is not necessarily a reflection group.
Still, the super Yang-Mills theory with a connected group $G$ is always a relative,
which is all that matters for our observation.

\subsection{$(SU(N)_k\times SU(N)_{-k})/\bZ_m$ ABJM theory}
Let us next study the $(SU(N)_k\times SU(N)_{-k})/\bZ_m$ ABJM theory,
where $m$ is a divisor of $N$.
This class includes the BLG theories as a special case when $N=2$.
For previous studies of the moduli space, see \cite{Lambert:2010ji,Bagger:2012jb}.

\paragraph{Consistency of the quotient:}
Here the quotient $\bZ_m$ is a subgroup of the diagonal subgroup of the center, $\bZ_N \subset \bZ_N\times \bZ_N$,  which does not act on the bifundamentals.
In general, the quotient introduces new topologically-nontrivial configurations of gauge fields,
and can make the Chern-Simons term with a given level ill-defined.
Therefore, as a zeroth step, we need to check that this quotient is consistent with the Chern-Simons level.
Equivalently, we need to check that the one-form  $\bZ_m$ symmetry we are trying to gauge is non-anomalous.
The general framework was given e.g.~in \cite{Gaiotto:2014kfa,Hsin:2018vcg};
in a more traditional language, the analysis can be presented as follows.

The Chern-Simons term is defined by extending the gauge field to an auxiliary 4d spacetime.
We therefore need to make sure that the value does not depend on the way we extend the gauge fields to 4d.
The condition to be checked is then \begin{equation}
\int_{M_4} k (\frac12\tr (\frac{F}{2\pi})^2 - \frac12\tr (\frac{F'}{2\pi})^2 ) \in \bZ
\label{foo}
\end{equation} for an arbitrary configuration of $(SU(N)\times SU(N))/\bZ_m$ gauge fields $(F,F')$ on a closed spin manifold $M_4$.

Now we note that the Stiefel-Whitney classes of the gauge fields satisfy $w_2(F)=w_2(F')= m'x \in \bZ_N$
where $x$ is an integer and $m'm=N$.
We also use the fact that the instanton number $\int_{M_4} \frac12\tr (\frac{F}{2\pi})^2$ modulo 1 is uniquely fixed by its Stiefel-Whitney class \cite{Witten:2000nv,Aharony:2013hda}.
Assuming this fact, it is convenient to compute this instanton number modulo 1 by taking a $U(1)$ configuration $\mathcal{F}$ and embedding it to $PSU(N)$ via \begin{equation}
\begin{aligned}
U(1) & \to PSU(N), \\
e^{\ii t} & \mapsto e^{\ii t \diag(1,1,\ldots,1-N)/N}.
\end{aligned}
\end{equation} We then find that \begin{equation}
\int_{M_4} \frac12\tr (\frac{F}{2\pi})^2 = 
-\frac{1}{N} \int_{M_4} \frac12 (\frac{\mathcal{F}}{2\pi})^2
\mod 1
\label{W}
\end{equation}
using the fact that \begin{equation}
\int_{M_4} \frac12 (\frac{\mathcal{F}}{2\pi})^2 \in \bZ 
\end{equation} on any spin manifold.
Then the relation \eqref{foo} immediately follows.

\paragraph{Determination of the moduli space:}

Let us study the moduli space. 
We give a generic vev to the bifundamentals, which is known \cite{Aharony:2008ug,Aharony:2008gk} to be given by $\bC^{4N}$ parametrized by  \begin{equation}
(z^I_1, z^I_2,\ldots, z^I_N) \label{z}
\end{equation} where $I=1,2,3,4$ are the $SU(4)_R$ indices, which we drop in the following. 

We now consider the subgroup $(S[U(1)^N] \times S[U(1)^N])/\bZ_m \rtimes S_N$ adapted to this generic vev.
The action of $S_N$ simply permutes $z_i$ and can be dealt with easily later.
Let us concentrate then on $H=(S[U(1)^N] \times S[U(1)^N])/\bZ_m$.
There is a subgroup $H'$ of $H$ which acts trivially on all the matter fields.
We  dualize the $H'$ gauge fields into  periodic scalars.
The group $H$ acts not only on the space $\bC^{4N}$ of $z_i$ but also on the periodic scalars, due to the Chern-Simons coupling.
The moduli space before identification by $S_N$ is then given by \begin{equation}
\frac{(\text{space $\bC^{4N}$ of $z_i$}) \times (\text{periodic scalars})}{H}
=\frac{(\text{space $\bC^{4N}$ of $z_i$})}{H''}.
\label{truemoduli}
\end{equation}
where $H''$ is the subgroup of $H$ fixing the periodic scalars.
To find $H''$ it is useful to note that the monopole operators of the subgroup $H'$ provides the functions parameterizing the periodic scalars.
This means that $H''$ is the subgroup of $H$ preserving all the monopole operators.

Let us now find this quotient.
We parametrize the group $H$  by \begin{equation}
(g_1^L,\ldots, g^L_{N}; g^R_1,\ldots, g^R_{N}) \label{g}
\end{equation} where $g^L_i$ and $g^R_i$ are complex numbers with absolute value $1$.
They act on $z_i$'s by the formula $z_i \mapsto (g^L_i/g^R_i)  z_i$.
We have an added constraint that \begin{equation}
\prod g^L_i = \prod g^R_i = 1,
\end{equation} 
and also need to impose the $\bZ_m$ identifications.

The monopole operators have the monopole charges \begin{equation}
(q_1,\ldots,q_N; q'_1,\ldots,q'_N)
\end{equation}
with the constraints \begin{equation}
\sum q_i=\sum q'_i=0,\qquad q_i = \frac{m'x_i}{N},\qquad q'_i = \frac{m'x'_i}{N} \quad\text{where $x_i, x'_i\in \bZ$ and $m'm=N$.}
\end{equation} 
We denote the corresponding monopole operator by $\cO_{(q_i,q'_i)}$.
The monopole charges form the $SU(N)$ root lattice when $m=1$ and the $SU(N)$ weight lattice when $m=N$, and the intermediate lattices when $m$ is in between. 
The element \eqref{g} acts on them by the formula
\begin{equation}
\cO_{(q_i,q'_i)}
\mapsto
\cO_{(q_i,q'_i)}
\prod_i (g^L_i)^{k q_i}/(g^R_i)^{k q'_i} .
\label{monopole-shifts}
\end{equation}
The element \eqref{g} obviously acts on $z_i$ via \begin{equation}
z_i \mapsto z_i (g^L_i/g^R_i), \qquad \text{(no summation on $i$)}.
\end{equation}
Below we focus on $h_i:=g^L_i/g^R_i$.

Let us study the case $m=N$ first. 
We see that the monopole whose charge is the $i$-th fundamental weight imposes the condition that 
$h_i$ is a $k$-th root of unity.
We also have an obvious condition that $\prod h_i=1$.
This means that identifications on $z_i$ are generated by \begin{equation}
(z_i,z_j) \mapsto (e^{2\pi\ii  /k } z_i, e^{-2\pi\ii /k} z_j), \qquad \text{others fixed.}
\end{equation}
Together with $S_N$, they form the group $G(k,k,N)$. 
When $N=2$, it is the dihedral group $\bD_{2k}$ with $2k$ elements.

Let us next consider the case $m=1$. The conditions imposed on $h_i$'s are now \begin{equation}
\text{$h_i/h_j$ is a $k$-th root of unity}
\end{equation} from the monopoles in the root lattice.
We also still have the condition $\prod_i h_i=1$. Like the previous case we still have  
the identification generated by $(h_i)=(e^{2\pi\ii /k},e^{-2\pi\ii /k},1,\ldots,1)$ and its permutations, but now we also have additional elements like $(h_i)=(e^{2\pi\ii /N},e^{2\pi\ii /N},\ldots,e^{2\pi\ii /N})$ and $(h_i)=(e^{\pi\ii /k},e^{-\pi\ii /k},e^{\pi\ii /k},\ldots,e^{-\pi\ii /k})$ for $N$ even.
Together with $S_N$, they \emph{do not} necessarily form a complex reflection group,
although they happen to do so for $N=2$, as we only have the added generator $(h_i)=(e^{\pi\ii /k},e^{-\pi\ii /k})$, which gives $\bD_{4k}$.

Our analysis reproduces the well-known results \cite{Distler:2008mk,Bagger:2012jb} for the $SU(2)_k\times SU(2)_{-k}$ and $(SU(2)_k\times SU(2)_{-k})/\bZ_2$ BLG theories.
The moduli spaces for the intermediate $\bZ_m$ subgroups of $\bZ_N$ can be similarly identified,
but we will not carry it out in detail here, since such cases (including the case $m=1$) all correspond to children of the $(SU(N)_k\times SU(N)_{-k})/\bZ_N$ theory.

\subsection{$(U(N+x)_k\times U(N)_{-k})/\bZ_p$ ABJ(M) theory}
Next, we consider the ABJ(M) theory of the form $(U(N+x)_k\times U(N)_{-k})/\bZ_p$,
where $\bZ_p$ is a subgroup of the diagonal $U(1)\subset U(1)_L\times U(1)_R$.
Let us first consider which $p$ is allowed.

\paragraph{Consistency of the quotient:}

It is clear that $p$ needs to be a divisor of $k$,
since the monopole operator introduced by a $\bZ_p$ quotient will have charge $k/p$ under $U(1)_L$,
which needs to be an integer.
Next, we need to study which $p$ is compatible with the Chern-Simons level $(k,-k)$.
This can be done as above, by studying the instanton number modulo 1 on the auxiliary 4-dimensional spin manifold.
We use the homomorphism \begin{equation}
\begin{aligned}
U(1)  & \to (U(N+x)_k \times U(N)_{-k})/\bZ_p, \\
e^{\ii t} & \mapsto (e^{\ii t \diag(1,1,\ldots,1)/p},e^{\ii t \diag(1,1,\ldots,1)/p})
\end{aligned}
\end{equation}
to embed a $U(1)$ configuration to $(U(N+x)_k\times U(N)_{-k})/\bZ_p$.
Then we find \begin{equation}
\int_{M_4} k(\frac12\tr (\frac{F}{2\pi})^2-\frac12\tr (\frac{F'}{2\pi})^2 )  =
( \frac{k (N+x) }{p^2 } - \frac{k N}{p^2}  )  \int_{M_4} \frac12 (\frac{\mathcal{F}}{2\pi})^2 \mod 1.
\end{equation}
We therefore need to require \begin{equation} 
\frac{k (N+x) }{p^2 } - \frac{k N}{p^2} = \frac{kx}{p^2}= \frac{\ell x}{p} \in \bZ
\label{consistency}
\end{equation} where $p\ell=k$.

The same result can be obtained in a slightly different but essentially in the same way.
The $\bZ_k$ 1-form symmetry we identified above is in general anomalous. 
The anomaly is measured by the topological spin of the line operator of charge $\ell\in \bZ_k$ representing the background for the 1-form symmetry.
The topological spin (which is defined modulo $1/2$ in a spin theory) can be computed easily to be \begin{equation}
\frac{(N+x)\ell^2}{2k} - \frac{N\ell^2}{2k} = \frac{\ell^2x}{2k} = \frac{\ell x}{2p}. \label{spin}
\end{equation} 
The anomaly-free lines are those for which the topological spin \eqref{spin} vanishes modulo $1/2$, reproducing \eqref{consistency}.

\paragraph{Determination of the moduli space:}
The moduli space can be found as above.
We consider a generic vev \eqref{z} of bifundamentals,
and study the adapted subgroup $H=(U(1)^{N} \times U(1)^N)/\bZ_p$.
The elements of $H$ are parameterized by \begin{equation}
(g_1^L,\ldots, g^L_{N}; g^R_1,\ldots, g^R_{N}) 
\end{equation} under the $\bZ_p$ identification.
The monopole operators have charges \begin{equation}
(q_1,\ldots,q_N;q'_1,\ldots, q'_N)
\end{equation} with the constraint that $q_i \equiv q/p $ modulo $1$, for some integer $q$, and similarly for $q'_i$.
An element of $H$ acts on a monopole operator of this charge $\cO_{(q_i,q'_i)}$ as before,
\begin{equation}
\cO_{(q_i,q'_i)}
\mapsto
\cO_{(q_i,q'_i)}
\prod_i (g^L_i)^{k q_i}/(g^R_i)^{k q'_i} .
\end{equation}

The requirement that all monopole operators $\cO_{(q_i,q'_i)}$ are fixed is therefore equivalent to the condition that i) $h_i:= g^L_i/g^R_i$ are $k$-th roots of unity and that ii) $(\prod h_i)^{k/p}=1$.
Its action on the bifundamental vevs $z_i$ is generated by \begin{equation}
(z_i,z_j) \mapsto (e^{2\pi\ii  /k } z_i, e^{-2\pi\ii /k} z_j), \qquad \text{others fixed,}
\end{equation} and \begin{equation}
z_i \mapsto e^{2\pi\ii  p/k} z_i, \qquad \text{others fixed.}
\end{equation}
Together with $S_N$, they form the group  $G(k,p,N)$.

In particular, we see that the moduli space of $(U(N)_k\times U(N)_{-k})/\bZ_k$ is always equal to the moduli space of $(SU(N)_k\times SU(N)_{-k})/\bZ_N$, both associated to the complex reflection group $G(k,k,N)$.
We will show in Sec.~\ref{sec:equivalence} that these two theories are in fact the same.

We also find that  the moduli space of $U(N)_k\times U(N)_{-k}$ is always a $\bZ_k$ quotient of the moduli space of $[SU(N)_k\times SU(N)_{-k}]/\bZ_N$.
We note that in the paper \cite{Lambert:2010ji} it was found that this statement was true only when $N$ and $k$ are coprime.
This is due to their additional condition that the $\bZ_k$ quotient should act diagonally on all $z_i$'s without any mixing with the gauge group, see the paragraph containing (59) in their paper.
Therefore our finding does not contradict theirs.

\subsection{$USp(2N)_k\times SO(2)_{-2k}$ type theories}

Next we consider the case with gauge group $USp(2N)_k\times SO(2)_{-2k}$. Generically, the family of theories with gauge group $USp(2N)_k\times SO(M)_{-2k}$ only have $\mathcal{N}= $ 5 SUSY, but for the special case of $M=2$, it is known to enhance to $\mathcal{N}= $ 6. We can then study the moduli space of this family of theories.

\paragraph{Consistency of the quotient:}
Here we can only quotient by a $\bZ_2$ subgroup whose generator is a combination of the $\bZ_2$ centers of $USp(2N)$ and $SO(2)$. Like in the previous cases, we first need to check that the quotient is consistent with the Chern-Simons level, which can be done using the same method as before. The $U(1)$ can be embedded as
\begin{equation}
\begin{aligned}
U(1)  & \to (USp(2N)\times SO(2))/\bZ_2, \\
e^{\ii t} & \mapsto (e^{\ii \frac{t}{2}\diag(1,-1,1,-1,\ldots,1,-1)},e^{\ii\frac{t}{2}\sigma_2}),
\end{aligned}
\end{equation}
where $\sigma_2$ stands for the appropriate Pauli matrix. We then find \begin{equation}
\int_{M_4} k(\frac12\tr (\frac{F}{2\pi})^2-\frac12\tr (\frac{F'}{2\pi})^2 )  =
( \frac{k N}{2} - \frac{k}{2}  )  \int_{M_4} \frac12 (\frac{\mathcal{F}}{2\pi})^2 \mod 1.
\end{equation}

We therefore need to require that $k(N-1)$ be even.

\paragraph{Determination of the moduli space:}
The moduli space can be found as in the previous cases.
We consider a generic vev \eqref{z} of bifundamentals, and study the remaining unbroken gauge group which in this case is rather simple. A generic bifundamental vev breaks the $USp(2N)_k\times SO(2)_{-2k}$ to $U(1)_{2k}\times U(1)_{-2k}$ and a decoupled $USp(2N-2)_{k}$ Chern-Simons theory\cite{Aharony:2008gk}. When we preform the $\bZ_2$ quotient these are changed to $(U(1)_{2k}\times U(1)_{-2k})/\bZ_2$ and $\frac{USp(2N-2)_{k}}{\bZ_2}$. Note that while we can always take the quotient in the former theory, the latter theory is only well defined if $k(N-1)$ is even, as expected from the previous analysis.

The moduli space is determined only by the $U(1)_{2k}\times U(1)_{-2k}$ theory. As discussed previously, the group one needs to take the quotient by is just $\bZ_{2k}$ in the $U(1)_{2k}\times U(1)_{-2k}$ case or $\bZ_{k}$ in the $(U(1)_{2k}\times U(1)_{-2k})/\bZ_2$ case. Both are complex reflection groups.

\subsection{More general variants of ABJ(M) theories}

All possible Lagrangian \Nequals6 theories were classified in \cite{Schnabl:2008wj} up to the level of the gauge algebras, not gauge groups\footnote{The classification only covers the cases where \Nequals6 supersymmetry is manifest in the Lagrangian. There could be cases of Lagrangian theories manifesting smaller supersymmetry, which enhances to \Nequals6 at low-energies. An example of this is given by the $USp(2N)_1\times O(M)_{-2}$ theories which classically have only \Nequals5, but are expected to have enhanced \Nequals6\cite{Aharony:2008gk}. Since this class of theories are expected to be dual to the $U(N+x)_4\times U(N)_{-4}$ family\cite{Aharony:2008gk,CGHNP} they do not give new moduli spaces.}.
Here we would like to analyze their moduli spaces. 

In \cite{Schnabl:2008wj}, it was shown that there are only two classes of Lagrangian \Nequals6 theories with unique energy momentum tensor.
The first is the ABJ(M) theories:
\begin{equation}
\mathfrak{g}=\mathfrak{su}(N+x)_k\times\mathfrak{su}(N)_{-k}\times \prod \mathfrak{u}(1)^n_{K_{ab}}  
\end{equation} 
with a bifundamental of $\mathfrak{su}\times\mathfrak{su}$ with charge $q_a$ under $a$-th $\mathfrak{u}(1)$, 
with the constraint \begin{equation}
\frac{1}k(\frac1{N+x}-\frac1{N}) = K^{ab} q_aq_b\label{n6condition}
\end{equation} where $K^{ab}$ is the inverse of the level matrix $K_{ab}$.
We can assume that $q_a$  are integers and that $\gcd(q_a)=1$ without loss of generality.
We assume $x\ge 0$.
This includes the case where $N=1$, so that one $\mathfrak{su}$ factor is actually missing.

The second is the theory
\begin{equation}
\mathfrak{g}=\mathfrak{usp}(2M)_k \times \mathfrak{u}(1)^n_{K_{ab}}  
\end{equation}
with a fundamental of $\mathfrak{usp}$ with charge $q_a$ under the $a$-th $\mathfrak{u}(1)$, with the constraint \begin{equation}
\frac{1}{2k} = K^{ab} q_aq_b.
\end{equation}
When $M=1$ this is a degenerate example of the ABJ(M) theory where $N+x=2$ and $N=1$ above.
We again assume that $q_a$  are integers and that $\gcd(q_a)=1$ without loss of generality.

We now need to worry about the global structure of the gauge group.
Denote by $G$ the connected Lie group corresponding to $\mathfrak{g}$ above,
chosen so that the simple part is simply-connected and that the abelian part was chosen so that $q_a$ are integers.
Then we can have a theory with the gauge group\begin{equation}
( G \rtimes X) / Z
\end{equation}
where $X$ is a finite group which might have a nontrivial outer automorphism action on $G$,
and $Z$ is a certain finite subgroup of $G\rtimes X$ compatible with the Chern-Simons levels.
The finite group part $X$ itself can have its Chern-Simons levels to make things more complicated.
Studying them all is a tiresome business.

Now we pick a particular relative. We first gauge the 0-form symmetry $\hat Z$ and arrive at the theory whose gauge group is $G\rtimes X$.
We then gauge the 1-form symmetry $\hat X$ to obtain the theory whose gauge group is just $G$.
We now pick a rather strange subgroup $Z'$ of the center of $G$ and consider $G/Z'$.
This needs to be chosen appropriately depending on various cases.

\subsubsection{ABJM theories}
Let us first consider the ABJM theories for which $G=SU(N)\times SU(N) \times U(1)^n$.
We pick the subgroup $Z'$ to be generated by two generators.
One is the diagonal combination \begin{equation}
(e^{2\pi\ii /N}, e^{-2\pi\ii /N})\in SU(N)\times SU(N)
\end{equation} and another is the combination \begin{equation}
\gamma:=(e^{2\pi\ii  m^a}) \in \prod_a U(1)^{(a)}.
\end{equation}

The conditions for $m^a$ are that i) they do not act on the hypers \begin{equation}
m^a q_a \in \mathbb{Z}, \label{1}
\end{equation} ii) the newly-introduced monopole operators should be integer-charged under $U(1)^n$: \begin{equation}
K_{ab} m^b \in \mathbb{Z},\label{2}
\end{equation} and iii) the Chern-Simons level is consistent with the quotient.
The third condition can be studied by using  \begin{equation}
\begin{aligned}
U(1) &\to U(1)^n/Z' \\
e^{i\theta} &\mapsto  e^{ i m^a \theta} 
\end{aligned}
\end{equation} to embed a $U(1)$ configuration $\mathcal{F}$ to $U(1)^n/Z'$.
As before, we find \begin{equation}
\int_{M_4} \frac{K_{ab}}2 \frac{F^{(a)}}{2\pi}\frac{F^{(b)}}{2\pi}
= K_{ab} m^a m^b \int_{M_4} \frac12 (\frac{\mathcal F}{2\pi})^2 
\label{K}
\end{equation} modulo 1. 
Therefore we find the condition \begin{equation}
K_{ab} m^a m^b \in \mathbb{Z}.\label{3}
\end{equation}
These three conditions \eqref{1}, \eqref{2} and \eqref{3} can be simultaneously solved by choosing \begin{equation}
m^a = K^{ab} q_a
\end{equation} thanks to \eqref{n6condition}.

To analyze the moduli space, we proceed as usual.
First, we give a generic vev to the hypers, and take an adapted subgroup $H$.
Second, we identify the subgroup $H'$ of $H$ which acts trivially on the matter fields.
Third, we enumerate all monopole operators of $H'$.
Fourth, we study the action of $H$ on the monopole operators, finding the subgroup $H''$ fixing them.
And finally, the moduli space is obtained by dividing $\bC^{4r}$ by $H''$.

In this case, the monopole operator $\cO_\gamma$ associated to the generator $\gamma$ has the charge $q^a$ under $U(1)^{(a)}$, and therefore has the same charge as the bifundamental.
Therefore, anything which fixes $\cO_\gamma$ fixes the bifundamental,
and can be forgotten as far as the moduli space is concerned.
Therefore the moduli space of this theory is the same as the moduli space of the theory $(SU(N)_k\times SU(N)_k)/\bZ_n$, which was already analyzed to be given by a complex reflection group.
It is also fairly clear that this theory is the `locally oldest' among the relatives.

\subsubsection{ABJ theories}
Next consider the case $G=SU(N+x)\times SU(N)\times U(1)^n$ with $x\neq 0$.
For this we take $Z'$ is to be generated by a single generator 
\begin{equation}
\gamma:=(e^{2\pi\ii /(N+x)}, e^{-2\pi\ii /N} , e^{2\pi\ii  m^a}  )\in SU(N+x)\times  SU(N) \times  \prod_a U(1)^{(a)}
\end{equation} where $m^a$ is chosen as follows.
First, this generator should act trivially on the matter fields: \begin{equation}
\frac{1}{N+x}-\frac{1}N + m^a q_a \in \mathbb{Z}. \label{1p}
\end{equation}
Second, we require that the monopole operator $\cO_\gamma$ is integer charged under $U(1)^{(a)}$
\begin{equation}
K_{ab}m^b \in \mathbb{Z} \label{2p}.
\end{equation}
 Finally, we need to ensure that the Chern-Simons interaction is consistent with the quotient.
 This can be studied as always by considering the embedding 
  \begin{equation}
 \begin{aligned}
 U(1)&\to (SU(N+x)\times SU(N)\times \prod_a U(1)^{(a)})/\vev{\gamma} ,\\
e^{\ii t} &\mapsto (
e^{\ii t\diag(-1+\frac1{N+x},\frac1{N+x},\cdots,-\frac1{N+x})},
e^{\ii t\diag(1-\frac1{N},-\frac1{N},\cdots,-\frac1{N})},
e^{\ii tm^a}
).
\end{aligned}\label{u1gamma}
\end{equation}
This gives the condition 
 \begin{equation}
k(-\frac{1}{N+x} + \frac{1}N) + K_{ab} m^a m^b \in \mathbb{Z},\label{3p}
\end{equation}
where we used previous calculations \eqref{W} and \eqref{K}.

The three conditions \eqref{1p}, \eqref{2p} and \eqref{3p} can be simultaneously solved by taking \begin{equation}
m^a = k K^{ab} q_b,
\end{equation}
thanks to the \Nequals6 condition \eqref{n6condition}.
Let us now study the moduli space of this theory.

The monopole operator $\cO_\gamma$ has charge $K_{ab} m^b = kq_a$ under $U(1)^{(a)}$,
which breaks it to $\bZ_{kq_a}$. 
Since the hypers have charge $q_a$ under $U(1)^{(a)}$, 
$\bZ_{kq_a}$ acts via $\bZ_k$ on the hypers, leading to the quotient \begin{equation}
(z_1,\ldots,z_n) \sim e^{2\pi\ii /k} (z_1,\ldots,z_n).
\end{equation}
$\cO_\gamma$ also has charge $k$ under the $U(1)$ subgroup of $SU(N+x)$ or $SU(N)$ given by \begin{equation}
\begin{aligned}
U(1) & \to SU(N) \\
e^{\ii t} & \mapsto e^{\ii t\diag(t,-t,1,\ldots, 1)}
\end{aligned}
\end{equation} and therefore we have the identification \begin{equation}
(z_1,\ldots,z_n) \sim (e^{2\pi\ii /k }z_1,e^{-2\pi\ii /k} z_2, z_3,\ldots,z_N ).\label{hoge}
\end{equation}
$\cO_\gamma$ itself has the charge 0 under the $U(1)_\gamma$ given in \eqref{u1gamma},
but $U(1)_\gamma$ acts trivially on $(z_1,\ldots,z_a)$.
Instead, let us consider the following subgroup $U(1)_{\gamma'}$ given by 
\begin{equation}
 \begin{aligned}
 U(1)&\to ( SU(N+x)\times SU(N)\times \prod_a U(1)^{(a)})/\vev{\gamma} \\
e^{\ii t} &\mapsto (
e^{\ii t\diag(\frac1{N+x},\frac1{N+x},\cdots,1-\frac1{N+x})},
e^{\ii t\diag(1-\frac1{N},-\frac1{N},\cdots,-\frac1{N})},
e^{\ii tm^a}
)
\end{aligned}.
\end{equation}
$\cO_\gamma$ has charge $k$ under $U(1)_{\gamma'}$, breaking it to $\bZ_k$.
This acts on the moduli space as \begin{equation}
(z_1,\ldots,z_N) \sim (e^{2\pi\ii /k }z_1, z_2, z_3,\ldots,z_N ).
\end{equation}
Together with the action of $S_N$ on the moduli space, they generate $G(k,1,N)$.

Now this particular choice $G/Z'$ is not guaranteed to be `locally oldest' among relatives.
One might try to take a quotient $G/Z''$ where $Z'\subset Z''$.
Even then, the identification by \eqref{hoge} remains. 
These identifications together with $S_N$ generate $G(k,k,N)$.
Therefore, any locally oldest relative above $G/Z'$ would give the group $\Gamma$ of identification which lies between the two extremes, \begin{equation}
G(k,k,N) \subset \Gamma \subset G(k,1,N).
\end{equation}
Such a $\Gamma$ is necessarily one of $G(k,x,N)$, from the following argument.
Recall \begin{equation}
\bC[z_1,\ldots,z_N]^{G(k,k,N)} = \bC[w_1,w_2,\cdots, w_{N-1}, w_N{=}e_N^k]
\end{equation} where $w_d$ is the elementary symmetric polynomial of degree $d$ constructed from $z_i^k$, and $e_N=z_1z_2\cdots z_N$.
Recall similarly \begin{equation}
\bC[z_1,\ldots,z_N]^{G(k,1,N)} = \bC[w_1,w_2,\cdots, w_{N-1}, e_N].
\end{equation}
Therefore \begin{equation}
\bC[z_1,\ldots,z_N]^{\Gamma} = \bC[w_1,w_2,\cdots, w_{N-1}, e_N^{p} ].
\end{equation} for some $p$, and hence $\Gamma$ is one of $G(k,x,N)$.

\subsubsection{$\mathfrak{usp}(2M)\times \mathfrak{u}(1)^N$ theories}
As far as the structure of the moduli space is concerned, the analysis for this last case was essentially already done above,
since the nonzero hypers can only be activated for a $\mathfrak{usp}(2)$ subgroup,
and this $m=1$ case happens to be included in the ABJ theory where $N+x=2$ and $N=1$. However, the decoupled $\mathfrak{usp}(2M-2)$ Chern-Simons theory still has an effect as it limits the existence of the $\bZ_2$ quotient.

\section{$(U(N)_k \times U(N)_{-k}) /\bZ_k = (SU(N)_k\times SU(N)_{-k})/\bZ_{N}$ }
\label{sec:equivalence}

In this section, we establish the equivalence of two ABJM theories based on 
$(U(N)_k \times U(N)_{-k}) /\bZ_k$ and $(SU(N)_k\times SU(N)_{-k})/\bZ_{N}$.
\subsection{An easier case}
Let us start with the case $N=1$ and $k=1$.
The statement in this case becomes the equivalence of two charged hypermultiplets and  $U(1)_1\times U(1)_{-1}$ coupled to the same hypermultiplets. 
As we will see, the equivalence still holds even when we replace the charged hypermultiplets with arbitrary system with a $U(1)$ symmetry.
Let us work in this generalized framework.
The fermions in the \Nequals3 Chern-Simons multiplets can be safely integrated away, 
therefore the statement to be shown is as follows.

Consider a theory with a $U(1)$ symmetry with the action $S[A_\mu]$.
We would like to establish that the theory with the action\begin{equation}
\pi\ii  \int \frac{a}{2\pi} d\frac{a}{2\pi}
-\pi\ii  \int \frac{b}{2\pi} d\frac{b}{2\pi}
+S[a-b]\label{a-b}
\end{equation} is equivalent to $S[0]$; here we follow Seiberg's convention that the lower-case fields are path-integrated over.

Note that Witten showed in \cite{Witten:2003ya} that the theory \begin{equation}
2\pi\ii  \int \frac{a}{2\pi} d\frac{b}{2\pi} + S[b]
\end{equation} is equivalent to $S[0]$. 
The point is that the integral over $a$ gives a delta function for $b$.
We can reduce our computation to Witten's case.
To see this,
we first rewrite \eqref{a-b} to \begin{equation}
\pi\ii  \int \frac{a+b}{2\pi} d\frac{a-b}{2\pi}
+S[a-b].
\end{equation} 
We are tempted to go to new variables by setting \begin{equation}
a':= a+b, \qquad b':= a-b.
\end{equation}
But there are two problems: the Chern-Simons levels differ by a factor of 2,
and the map from $(a,b)$ to $(a',b')$ cannot be inverted.

Instead, we simply  use the variables $a$ and $b'=a-b$.
Then we have \begin{equation}
\pi\ii  \int \frac{a}{2\pi} d\frac{a}{2\pi}
-\pi\ii  \int \frac{b}{2\pi} d\frac{b}{2\pi}
+S[a-b]
= 2\pi\ii  \int \frac{a}{2\pi} d\frac{b'}{2\pi}
-\pi\ii  \int \frac{b'}{2\pi} d\frac{b'}{2\pi}
+S[b'].
\end{equation}
Now the integral over $a$ gives the delta function for $b'$, and we are done.

For generalization, it is useful to view the computation in the following way.
We first note that the action \eqref{a-b} has a one-form  $U(1)$ symmetry \begin{equation}
(a,b) \mapsto (a+c, b+c)
\end{equation} where $c$ is another $U(1)$ gauge field.
Then, we can perform the path integral $[DaDb]$ in two steps: \begin{enumerate}
\item We integrate along the direction of $U(1)$ one-form symmetry. 
At this point, we have an action functional depending on $b':=a-b$  parametrizing the orbits of the action of the one-form symmetry.
\item We then integrate along the $b'$ direction.
\end{enumerate}
There is usually not a very natural way to parametrize the direction of the $U(1)$ one-form symmetry,
but any choice would do. 
Here we just used $a$, but we can parametrize it by fixing a representative $(a_0,b_0)$ and 
integrating over the direction $(a_0+c,b_0+c)$.
Then we perform the path integral over $c$ of the form \begin{equation}
\int [Dc] \exp({2\pi\ii  \int \frac{c}{2\pi} d (\frac{a}{2\pi}-\frac{b}{2\pi}) + \cdots})
\end{equation}
which produces the delta function equating the $U(1)$ bundle $a$ and $b$.

\subsection{The general case}

Let us proceed to the general case.
The $(U(N)_k\times U(N)_{-k})/\bZ_k$ theory has a $U(1)$ one-form symmetry,
given by the following embedding \begin{equation}
0\to \underline{U(1)} \to (U(N)\times U(N))/\bZ_k \to   (U(N)\times U(N))/U(1)\to 0
\end{equation}
where $e^{\ii t}$ is sent to $(e^{\ii t/k},e^{\ii t/k})$ by the second arrow,
and the underline was added to distinguish various different $U(1)$s.

We perform the path integral over the direction of this $U(1)$ one-form symmetry,
then we obtain a functional of $(U(N)\times U(N))/U(1)$ bundles.
We need to understand how this can be related to $(SU(N)\times SU(N))/\bZ_N$ bundles.
The answer is that we have another sequence \begin{equation}
0\to (SU(N)\times SU(N))/\bZ_N \to   (U(N)\times U(N))/U(1) \to \doubleunderline{U(1)} \to 0
\end{equation}
where the third arrow sends $(g_1,g_2)$ to $\det g_1/\det g_2$.

Therefore, what we want to achieve is to first integrate over the $\underline{U(1)}$ direction
to generate the delta function for $\doubleunderline{U(1)}$, and we are done.
In this more general case, it is hard to find a nice change of variables to isolate the $\underline{U(1)}$ direction, but we can parameterize the $\underline{U(1)}$ direction by a $U(1)$ gauge field $c$.
Then the relevant part of the action is \begin{equation}
2\pi\ii  \int \frac{c}{2\pi} d(\tr\frac{F_1}{2\pi}-\tr \frac{F_2}{2\pi}) 
\end{equation} thanks to the cancellation of the level $k$ and $\bZ_k$ quotient.
The integration over $c$ provides the desired delta function for $\doubleunderline{U(1)}$, setting the determinants of two $U(N)$ bundles  to be equal.
The gauginos for $\underline{U(1)}$ and $\doubleunderline{U(1)}$ pair up and can be trivially integrated out.
This establishes the desired equivalence of the two ABJM theories.

\subsection{Comparing superconformal indices}

The preceding discussion might have sounded somewhat abstract to the reader.
As an additional check, let us directly compare the superconformal indices\cite{BBMR,SKim,ImYo} of the two theories. 

First, consider the index for the $(U(N)_k\times U(N)_{-k})/\bZ_k$ theory:
\begin{multline}
I = \sum_{m_i, n_i} \frac{1}{(N!)^2} \int \prod^{N}_{i=1} (\frac{z^{k m_i}_i d z_i}{2\pi\ii  z_i})(\frac{\tilde{z}^{-k n_i}_i d \tilde{z}_i}{2\pi\ii  \tilde{z}_i}) \frac{1}{\nu^{\sum_i k m_i}_1} \frac{1}{\nu^{\sum_i k n_i}_2} \\
\times x^{\sum_{i,j} 2 (|m_i - n_j| - |m_i - m_j| - |n_i - n_j|)} I_{PE} ({z_i}, {\tilde{z}_i}, {m_i}, {n_i})
\end{multline}

Here we use $z_i$ for the gauge fugacities of $U(N)_k$ and $\tilde{z_i}$ for those of $U(N)_{-k}$. The sum is over the monopole lattice of the two groups, spanned by $(m_1, m_2, ... , m_N)$ and $(n_1, n_2, ... , n_N)$ for $U(N)_k$ and $U(N)_{-k}$, respectively. We also use $\frac{1}{\nu^k_1}$ and $\frac{1}{\nu^k_2}$ for the fugacities for the topological symmetries of the two theories, where the non-standard definition is done for future use. The term $I_{PE} ({z_i}, {\tilde{z}_i}, {m_i}, {n_i})$ contains the plethystic exponential of the one particle index whose exact form will not be of use to us.

First let us separate the two $U(1)$ groups, for which it is convenient to first consider the case of $U(N)_k\times U(N)_{-k}$ and later take the $\bZ_k$ quotient. The basic monopole of $U(N)$ is of the form $(1,0,...,0)$ from which the rest can be generated by additions, reflections and permutations. In terms of $U(1)$ and $su(N)$ monopoles, it can be written as $(1,0,...,0) = (\frac{1}{N},\frac{1}{N},...,\frac{1}{N}) + (\frac{N-1}{N},-\frac{1}{N},...,-\frac{1}{N})$. We therefore see that the non-abelian part should be taken to be $[SU(N)_k\times SU(N)_{-k}]/\bZ_N$, and the monopole sum should ran over monopoles of the form $m (\frac{1}{N},\frac{1}{N},...,\frac{1}{N})$.

We next define $z_i = r z^{su}_i$  and $\tilde{z}_i = \tilde{r} \tilde{z}^{su}_i$ with $\prod z^{su}_i=\prod \tilde{z}^{su}_i=1$. One can show that:
\begin{multline}
\prod^{N}_{i=1} (\frac{z^{k m_i}_i d z_i}{2\pi\ii  z_i})(\frac{\tilde{z}^{-k n_i}_i d \tilde{z}_i}{2\pi\ii  \tilde{z}_i}) = \\
  \frac{N^2 r^{k \sum m_i}  \tilde{r}^{-k \sum n_i} dr d\tilde{r} \prod^{N-1}_{i=1} (z^{su}_i)^{k(m_1-m_N)} (\tilde{z}^{su}_i)^{-k(n_1-n_N)} dz^{su}_i d\tilde{z}^{su}_i}{(2\pi\ii  r) (2\pi\ii  \tilde{r}) \prod^{N-1}_{i=1} (2\pi\ii  z^{su}_i) (2\pi\ii  \tilde{z}^{su}_i)} .
\end{multline}

Next, we perform the integration over the non-abelian part and perform the summation over the non-abelian monopoles. The result should be the index for $[SU(N)_k\times SU(N)_{-k}]/\bZ_N$, which we shall denote as $I_{SU}$, evaluated in the presence of background abelian monopoles of the form $m (\frac{1}{N},\frac{1}{N},...,\frac{1}{N})$ and $n (\frac{1}{N},\frac{1}{N},...,\frac{1}{N})$ for the two $U(1)$ groups, which appear as global symmetries of the $[SU(N)_k\times SU(N)_{-k}]/\bZ_N$ theory. Note that since all matter fields are bifundamentals, the diagonal combination of the two $U(1)$ groups acts trivially. As a result, $I_{SU}$ depends only on $|m-n|$ and $\frac{r}{\tilde{r}}$ and not on $|m+n|$ and $r\tilde{r}$. The index of the $(U(N)_k\times U(N)_{-k})/\bZ_k$ theory could then be written as:
\be
I = \sum_{m, n} \int \frac{r^{k m}  \tilde{r}^{-k n} dr d\tilde{r}}{(2\pi\ii  r) (2\pi\ii  \tilde{r})} \frac{1}{\nu^{k m}_1}\frac{1}{\nu^{k n}_2} x^{2N|m - n|} I_{SU} (r, \tilde{r}, |m-n|).
\ee

Because of the $\bZ_k$ quotient, the monopole charges are quantized in units of $\frac{1}{k}$. It is convenient to redefine $m\rightarrow \frac{m}{k}$ and $n\rightarrow \frac{n}{k}$ so that the sum is over integer charges. We shall also make one final change of variables to $u= r\tilde{r}$ and $v= \frac{r}{\tilde{r}}$. We then have: 
\be
I = \sum_{m, n} \int \frac{u^{\frac{m-n}{2}}  v^{\frac{m+n}{2}} du dv}{(2\pi\ii  u) (2\pi\ii  v)} \frac{1}{\nu^{m}_1}\frac{1}{\nu^{n}_2} x^{\frac{2N}{k}|m - n|} I_{SU} (v, |m-n|).
\ee
 
Particularly, $I_{SU}$ is independent of $u$. We can then perform the integration over $u$, which has the effect of a Delta function setting $m=n$. This simplifies the expression to:  
\be
I = \sum_{m} \int \frac{v^m}{2\pi\ii  v dv} \frac{1}{(\nu_1 \nu_2)^{m}} I_{SU} (v, 0).
\ee

We can next expand $I_{SU} (v, 0)$ in a power series in $v$:
\be
I_{SU} (v, 0) = \sum_i I_i v^i. 
\ee

Inserting this in the expression of the index, we can now perform the integration over $v$, which has the effect of a $\delta$ function setting $m=-i$. Finally we get:
 \be
I = \sum_{i} (\nu_1 \nu_2)^{i} I_i = I_{SU} (\nu_1 \nu_2, 0).
\ee
Thus, we see that the two indices match with the baryonic symmetry on the $SU$ side being mapped to the diagonal topological symmetry on the $U$ side.

Finally we note that in \cite{Honda:2012ik} the agreement of the superconformal indices of the two theories was studied in the context of the equivalence of the $U(N)_k\times U(N)_{-k}$ theory 
and the $(SU(N)_k\times SU(N)_k)/\bZ_N$ theory further gauged by a $\bZ_k$ subgroup of the baryonic symmetry.
They found the agreement  only when $k$ and $N$ are coprime.
This was due to their assumption that the $\bZ_k$ part acts diagonally as a subgroup of the $U(1)$ baryonic symmetry, without mixing with the gauge group.
Our $\bZ_k$ action is more general and therefore our result does not contradict theirs.

\section*{Acknowledgments}

The authors thank Dongmin Gang for the collaboration at the early stage of the work. The authors also wish to thank Oren Bergman and Shlomo Razamat for useful discussions.
YT and GZ are supported in part by World Premier International Research Center Initiative (WPI), MEXT, Japan.
YT is also partially supported by JSPS KAKENHI Grant-in-Aid (Wakate-A), No.17H04837
and JSPS KAKENHI Grant-in-Aid (Kiban-S), No.16H06335.

\appendix
\section{Complex reflection groups}
\label{sec:app}
Here we give a short summary of complex reflection groups (which are also called unitary reflection groups).
The following is a very brief summary of \cite{LehrerTaylor,Springer}, in which much more details can be found.
The review articles \cite{GeckMalle,Dolgachev} are also a good resource.
Complex reflection groups appeared in the mathematical physics literature previously in e.g.~\cite{Cecotti:2015hca,Aharony:2016kai,Caorsi:2018zsq,Bonetti:2018fqz}.

\subsection{Definitions}
A pseudoreflection, or simply a reflection, on $\bC^n$ is a unitary transformation which acts on a one-dimensional subspace by a multiplication by a root of unity and fixes the orthogonal $(n-1)$-dimensional subspace.
A complex reflection group $G$ is a finite group generated by pseudoreflections, acting on $V\simeq \bC^n$.
The dimension $n$ is known as the rank of $G$.
A complex reflection group $G$ is called irreducible if $V$ is an irreducible representation of $G$.
A  complex reflection group is crystallographic if it preserves a lattice $\mathbb{Z}^{2n} \subset V\simeq \bC^{n}$.

A real reflection group, which might be better known, is defined by replacing $\bC$ by $\bR$ and pseudoreflections by ordinary reflections.
A real reflection group gives a complex reflection group by complexification.

Precisely speaking, a reflection group refers to the pair $(G,V)$. 
As an abstract group, $G$ has many representation on which it does not act as a reflection group.

\subsection{The Chevalley-Shephard-Todd theorem}

The theorem of Chevalley-Shephard-Todd says that,
given a complex linear space $V\simeq \bC^n$ acted on by a finite group $G$,
the invariant ring $\bC[V]^G$ is a free polynomial ring if and only if $G$ is a complex reflection group.
Here we assume $G$ is a subgroup of the unitary group without loss of generality, and identify $V\simeq V^*$. 

The proofs are ingenuous but not so difficult, and the proof of the only if part in particular would be of some interest to those familiar with the superconformal indices. 
So let us reproduce them here.
Along the way, we also obtain two important relations satisfied by the degrees of invariants.

We first introduce some notations.
Let $\bC[V]^G_+$ be
the subring of invariant polynomials with zero constant term,
and $\vev{\bC[V]^G_+}$ be  the ideal of $\bC[V]$ generated by $\bC[V]^G_+$.
For a polynomial $P\in \bC[V]$, we define $\Av{P}$  to be \begin{equation}
\Av{P}=\frac{1}{|G|}\sum gP \in \bC[V]^G.
\end{equation}
We also need a small fact: 
Pick a pseudoreflection $r\in G$ associated to a hyperplane $H\in V$. 
 Then, for any homogeneous polynomial $P\in \bC[V]$,
there is a polynomial $Q$ such that $(r-1)P=HQ$, where $\deg Q<\deg P$.
We start by proving a lemma:
\paragraph{Lemma.}
Let $U_1,\ldots, U_r$ be homogeneous elements of $\bC[V]^G$
and $P_1,\ldots, P_r$ be homogeneous elements of $\bC[V]$.
Suppose $U_1$ is not in the ideal generated by $U_2$, \ldots, $U_r$, and \begin{equation}
P_1U_1+\cdots + P_r U_r =0.
\end{equation}
Then $U_1 \in \vev{\bC[V]^G_+}$.
\paragraph{Proof of the lemma.}
We proceed by the induction in the degree of $P_1$.
If $P_1$ is a nonzero constant, then $\Av{P_1}=P_1$. 
We also have $\Av{P_1}U_1+\cdots + \Av{P_r} U_r=0$.
Therefore $U_1$ is in the ideal generated by $U_2$ to $U_r$, contradicting our assumption.

So let us assume $P_1$ has a nonzero degree.
Pick a pseudoreflection $r$ and write $(r-1)P_i=HQ_i$.
We then have $Q_1U_1+\cdots+Q_r U_r=0$.
By induction we know $Q_1\in \vev{\bC[V]^G_+}$.
Therefore $rP_1-P_1 = HQ_1 \in  \vev{\bC[V]^G_+}$.
Since $G$ is generated by pseudoreflections, we then have \begin{equation}
gP_1-P_1\in \vev{\bC[V]^G_+} 
\end{equation}for arbitrary $g$, and therefore \begin{equation}
\Av{P_1}-P_1\in \vev{\bC[V]^G_+},
\end{equation} and therefore $P_1$ itself is in $\vev{\bC[V]^G_+}$.

\paragraph{Proof of the if part.}
Let $I_1$, \ldots, $I_r$ be a minimal set of homogeneous generators of $\bC[V]^G$, and assume that they have a nontrivial homogeneous relation of the form \begin{equation}
H(I_1,\ldots,I_r)=0
\end{equation} for a polynomial $H(Y_1,\ldots, Y_r)$.
Let $H_i := \partial H/\partial Y_i$ and $U_i:=H_i(I_1,\ldots,I_r)$.
We can relabel the indices so that $U_1,\ldots,U_s$ generate all $U_1,\ldots,U_r$,
and that $s$ is a minimal such choice. Write \begin{equation}
U_k = \sum_{j=1}^s V_{kj} U_j
\end{equation} for $k>r$. 
We denote the coordinates of $V$ as $X_1$ to $X_n$. We have \begin{equation}
0=\frac{\partial H}{\partial X_i} 
=\sum_{j=1}^s U_j ( \frac{\partial I_j}{\partial X_i}+ \sum_{k=s+1}^r V_{kj} \frac{\partial I_k}{\partial X_i}).
\end{equation}
From the lemma, we see \begin{equation}
\frac{\partial I_j}{\partial X_i}+ \sum_{k=s+1}^r V_{kj} \frac{\partial I_k}{\partial X_i}\in \vev{\bC[V]^G_+}
\end{equation} for all $i$ and $1\le j\le s$, and therefore there are polynomials $B_{ij\ell}$ such that \begin{equation}
\frac{\partial I_j}{\partial X_i}+ \sum_{k=s+1}^r V_{kj} \frac{\partial I_k}{\partial X_i} = \sum_{\ell=1}^r B_{ij\ell} I_\ell,
\end{equation} where in the sum on the right hand side, only $I_\ell$ whose degree is lower than $I_j$ can appear. In particular, $B_{ijj}=0$.

We now use the Euler's formula that $\sum_i X_i \partial I_j/\partial X_i = d_j X_j$ where $d_j=\deg I_j$.
We find \begin{equation}
(\deg I_j) I_j + \sum_{k=s+1}^r V_{kj} (\deg I_k) I_k = \sum_{\ell=1}^r \Av{\sum_i (X_i B_{ij\ell})} I_\ell,
\end{equation} and the coefficient $\Av{\sum_i X_i B_{ijj}}$ multiplying $I_j$ on the right hand side is zero. 
Therefore $I_j$ is in the ideal generated by $I_{k\neq j}$, and contradicts our minimality assumption.

\paragraph{On the degrees of invariants.}
Let us say $d_1\le d_2 \le \cdots \le d_n$ be the degree of invariants.
The `unrefined index' $P(t)=\sum (\dim (\bC[V]^G)_n) t^n $ can be written in two ways:
\begin{equation}
\prod_i \frac{1}{1-t^{d_i}} = \frac{1}{|G|} \sum_g \frac{1}{\det_V (1-gt)}; \label{molien}
\end{equation}
this is Molien's theorem.

Now, multiply $(1-t)^n$ on both sides and take the $t\to 1$ limit. 
Only the term $g=e$ survives on the right hand side, and we obtain \begin{equation}
\prod d_i = |G|.
\end{equation}
Next, subtract $1/(1-t)^n$ on both sides, multiply by $(1-t)^{n-1}$ on both sides, and then take the $t\to 1$ limit.
Only the pseudoreflections contribute on the right hand side, and we obtain \begin{equation}
\frac12\sum (d_i-1)=\sum_{r\in R} \frac{1}{1-\lambda(r)}
\end{equation} where $R\subset G$ is the subset of pseudoreflections, and $\lambda(r)$ is the unique non-one eigenvalue of $r$.
Combining contributions from pseudoreflections associated to a single hyperplane, we obtain \begin{equation}
\sum (d_i-1)= |R|.
\end{equation}

\paragraph{Proof of the only if part.}
Suppose $\bC[V]^G$ is a free polynomial algebra with generators $I_1$, \ldots, $I_n$
with degrees $d_1\le d_2 \le \cdots \le d_n$.
The discussion in the previous subsection can be carried out without change,
and we find in particular \begin{equation}
\prod d_i=|G|, \qquad \sum (d_i-1)= |R|,
\end{equation}
where $R\subset G$ is the subset of pseudoreflections in $G$.
Let $G_0\subset G$ be the subgroup generated by $R$.
From the if part, $\bC[V]^{G_0}$ is a free polynomial algebra with generators $J_1,\ldots,J_n$,
with degrees $e_1\le e_2\le \cdots \le e_n$, 
and  in particular $\sum (e_i-1)=|R|$.
Since $\bC[V]^G \subset \bC[V]^{G_0}$,
a standard argument shows that $e_1 \le d_1$, $e_2\le d_2$, \ldots, $e_n\le d_n$.
Therefore $|R|=\sum (e_i-1)\le \sum (d_i-1) = |R|$,
and $d_i=e_i$.
Therefore, $|G_0|=\prod e_i = \prod d_i=|G|$,
meaning that $G_0=G$,
meaning that $G$ is generated by its reflections.

\subsection{Classification}
\paragraph{Real reflection groups:}
The list of real reflection groups is well-known: it consists of Weyl groups\footnote{%
We used a different notation in the main text, where the Weyl group of a Lie group $G$ was denoted by $\mathcal{W}_G$, and $\mathcal{H}_{3,4}$ by $H_{3,4}$.
} \begin{equation}
\mathcal{A}_n,\quad \mathcal{B}_n=\mathcal{C}_n,\quad \mathcal{D}_n,\quad \mathcal{E}_{6,7,8},\quad \mathcal{F}_4,\quad \mathcal{G}_2
\end{equation} of the corresponding root lattices, together with \begin{equation}
I_2(m), \quad \mathcal{H}_{3,4}.
\end{equation} 
Here, $I_2(m)$ is the dihedral group $\bD_{2m}=\bZ_m \rtimes \bZ_2$,
with $I_2(3)=\mathcal{A}_2$, $I_2(4)=\mathcal{B}_2$, $I_2(6)=\mathcal{G}_2$. 
Finally, $\mathcal{H}_3$ is the symmetry group of the dodecahedron (or its dual, the icosahedron) in $\bR^3$,
and $\mathcal{H}_4$ is the symmetry group of the 120-cell (or its dual, the 600-cell) in $\bR^4$.
In passing, we mention that $\mathcal{F}_4$ is the symmetry group of the 24-cell.

\paragraph{Complex reflection groups in an infinite series:}
The complex reflection groups which are not real reflection groups consist of a single infinite family and 28 exceptions.
The groups in the infinite families are denoted by $G(m,p,n)$, acting on $\bC^n$,
where $p|m$.
They are generated by $S_n$ acting on $z_{i=1,\ldots, n}$,
together with \begin{equation}
z_i \mapsto \gamma^p z_i,\qquad \text{other $z_j$ fixed},
\end{equation} and \begin{equation}
(z_i,z_j) \mapsto (\gamma z_i, \gamma^{-1} z_j), \qquad \text{other $z_k$ fixed},
\end{equation} where $\gamma=\exp(2\pi\ii /m)$.

We note that $G(m,p,1)$ is simply $\bZ_{m/p}$, and the following are real reflection groups:
\begin{itemize}
\item $G(1,1,n)$ is the symmetric group acting on $\bC^n$. This action is reducible.
\item $G(2,2,n)$ is the Weyl group  $\mathcal{D}_n$. 
\item $G(2,1,n)$ is the Weyl group  $\mathcal{B}_n=\mathcal{C}_n$. 
\item $G(m,m,2)$ is the dihedral group $I_2(m)$ of order $2m$.
\end{itemize}

The invariants are symmetric polynomials of $z_i^m$, together with $(z_1\cdots z_n)^{m/p}$.
Therefore the degree of invariants are: \begin{equation}
m, 2m, \ldots, (n-1)m; n(m/p).
\end{equation}

\paragraph{Exceptional complex reflection groups:}
The 28 exceptional complex reflection groups, together with the exceptional real reflection groups $\mathcal{E}_{6,7,8}$, $\mathcal{F}_4$, $\mathcal{H}_{3,4}$ are often labeled as $G_{4}$ to $G_{37}$, following Shephard and Todd.

The classification of exceptional reflection groups is usually done in two steps,
first by considering those with rank $=2$, and then by considering those with rank $\ge 3$.
This reflects the fact that the projection map $U(n)\to U(n)/U(1)$, if restricted to the subset of pseudoreflections, is 1:1 when $n\ge  3$ but is 2:1 when $n=2$.

Rank-two reflection groups are finite subgroups of $U(2)$.
As such, each of them has a finite subgroup of $U(2)/U(1)=SO(3)$ as a quotient.
The exceptional ones therefore correspond to the tetrahedral, octahedral or icosahedral group $\subset SO(3)$.
Those corresponding to the tetrahedral group are $G_4$ to $G_7$;
those  corresponding to the octahedral group are $G_8$ to $G_{15}$;
those  corresponding to the icosahedral group are $G_{16}$ to $G_{22}$.

Exceptional reflection groups with rank $\ge 3$ can be conveniently labeled by the set of the one-dimensional eigenspaces of pseudoreflections; this is a generalization of the concept of the set of roots of a Weyl group, and is known as the corresponding line system.
The order of pseudoreflections is at most three.
There is a single group $G_{26}=\mathcal{M}_3$ which contains reflections of order 2 and order 3.
There are two groups $G_{25}=\mathcal{L}_3$ and $G_{32}=\mathcal{L}_4$ which contains only reflections of order 3.
All the other complex reflection groups of rank $\ge 3$ contains only reflections of order 2.
Among them, those which do not come from real reflection groups are: 
two rank-3 groups, $G_{24}=\mathcal{J}_3^{(4)}$ and $G_{27}=\mathcal{J}_3^{(5)}$;
two rank-4 groups, $G_{29}=\mathcal{N}_4$ and $G_{31}=\mathcal{O}_4$;
a single rank-5 group $G_{33}=\mathcal{K}_5$ and a single rank-6 group $G_{34}=\mathcal{K}_6$.
The information are gathered in Table~\ref{table}.

\begin{table}
\vbox{}
\vspace*{-1cm}
\[
\begin{array}{cc|cc|cccc|l}
&\text{rank}&\text{ST name}&\text{type}&2 & 3 & 4 & 5 & \text{degrees of invariants}\\
\hline
\hline
\checkmark&2&G_4 &(\mathcal{T}) & & 4 && &4,6\\
\checkmark&2&G_5 &(\mathcal{T})& &4+4 &&&6,12\\
&2&G_6 &(\mathcal{T})& 6 & 4 &&&4,12\\
&2&G_7 &(\mathcal{T})&6& 4+4 &&&12,12\\
\hline
\checkmark&2&G_8 &(\mathcal{O})&6&&6&&8,12\\
&2&G_9 &(\mathcal{O})&12+6 & & 6 &&8,24\\
&2&G_{10} &(\mathcal{O})& 6 & 8 & 6 & &12,24\\
&2&G_{11}&(\mathcal{O})&12+6 & 8 & 6 &&24,24\\
\checkmark&2&G_{12}&(\mathcal{O})&12 &&&&6,8\\
&2&G_{13}&(\mathcal{O})&12+6 &&&&8,12\\
&2&G_{14}&(\mathcal{O})&12& 8 &&&6,24\\
&2&G_{15}&(\mathcal{O})&12+6& 8 &&&12,24\\
\hline
&2&G_{16}&(\mathcal{I})&&&&12&20,30\\
&2&G_{17}&(\mathcal{I})&30&&&12&20,60\\
&2&G_{18}&(\mathcal{I})&&20&&12&30,60\\
&2&G_{19}&(\mathcal{I})&30&20&&12&60,60\\
&2&G_{20}&(\mathcal{I})&&20&&&12,30\\
&2&G_{21}&(\mathcal{I})&30&20&&&12,60\\
&2&G_{22}&(\mathcal{I})&30&&&&12,20\\
\hline\hline
\rowcolor{lightgray}
&3&G_{23}&\mathcal{H}_3&15 &&&& 2,6,10\\
\checkmark&3&G_{24}&\mathcal{J}_3^{(4)}&21 &&&& 4,6,14\\
\checkmark&3&G_{25}&\mathcal{L}_3&& 12 &&& 6,9,12\\
\checkmark&3&G_{26}&\mathcal{M}_3&9&12 &&& 6,12,18\\
&3&G_{27}&\mathcal{J}_3^{(5)}&45 &&&& 6,12,30\\
\hline\hline
\rowcolor{lightgray}
\checkmark&4&G_{28}&\mathcal{F}_4&12+12 &&&& 2,6,8,12\\
\checkmark&4&G_{29}&\mathcal{N}_4&40 &&&& 4,8,12,20\\
\rowcolor{lightgray}
&4&G_{30}&\mathcal{H}_4&60 &&&& 2,12,20,30\\
\checkmark&4&G_{31}&\mathcal{O}_4&60 &&&& 8,12,20,24\\
\checkmark&4&G_{32}&\mathcal{L}_4& &40&&& 12,18,24,30\\
\hline\hline
\checkmark&5&G_{33}&\mathcal{K}_5&45 &&&& 4,6,10,12,18\\
\hline\hline
\checkmark&6&G_{34}&\mathcal{K}_6&126 &&&& 6,12,18,24,30,42\\
\rowcolor{lightgray}
\checkmark&6&G_{35}&\mathcal{E}_6&36&&&& 2,5,6,8,9,12\\
\hline\hline
\rowcolor{lightgray}
\checkmark&7&G_{36}&\mathcal{E}_7&63 &&&& 2,6,8,10,12,14,18\\
\hline\hline
\rowcolor{lightgray}
\checkmark&8&G_{37}&\mathcal{E}_8&120 &&&& 2,8,12,14,18,20,24,30
\end{array}
\]
\caption{Data of exceptional complex reflection groups.
The columns are for the Shephard-Todd name,
the corresponding subgroup of $SO(3)$ when  rank $=2$
or  the corresponding line system when rank $\ge 3$,
the number of pseudoreflections in each order (where $+$ denotes the existence of more than one conjugacy classes),
and the degrees of invariants.
Real reflection groups are shaded,
and the ones with $\checkmark$ are crystallographic.
\label{table}}
\end{table}

\paragraph{Crystallographic reflection groups.}
As already stated, a reflection group $G$ acting on $V\simeq \bC^n$ is called crystallographic if it preserves a lattice $\mathbb{Z}^{2n} \subset V$.
An equivalent condition is that the representation matrices of $G$ on $V$ can be defined in a imaginary quadratic extension $\mathbb{Q}[\sqrt{-d}]$.
 Real crystallographic reflection groups are also known as Weyl groups.
 Among complex reflection groups $G(m,p,n)$,
 the crystallographic ones are those with $m=2,3,4,6$.
 Among the exceptionals, the crystallographic ones are denoted by $\checkmark$
 in Table~\ref{table}.
 More details can be found in the review~\cite{Dolgachev} and references therein.

\subsection{On the invariants of rank-2 reflection groups}
Here we would like to describe the invariants of rank-2 reflection groups,
which can be understood in a systematic manner \cite{Springer}.
Let $G$ be a rank-2 complex reflection group,
and $\Gamma$ be a corresponding binary polyhedral group.
(We exclude cyclic groups of odd order in the analysis below.)
Let both of them act on $V\simeq \bC^2$.

We first consider semi-invariants of $\Gamma$, namely 
polynomials $P\in \bC[V]$ such that $gP=c(g)P$ for a homomorphism $c:\Gamma\to U(1)$.
One way to construct semi-invariants is the following.
$\Gamma$ acts also on $\CP^1\simeq S^2$.
Given a point $a=[a_1:a_2]\in \CP^1$,
we denote by $f_a$ the linear function $a_1 X_1+a_2 X_2 \in V^*$,
where $X_{1,2}$ are two coordinates of $V$.
$f_a$ is well-defined up to a scalar multiple.
Pick a $\Gamma$-orbit $O$ in $\CP^1\simeq S^2$,
and let $f_O := \prod_{a\in O} f_a$.
This is clearly gives a semi-invariant.

Conversely, given a homogeneous semi-invariant, its zeroes clearly determine a $\Gamma$-invariant divisor (i.e.~a formal  integer linear combination of $\Gamma$-orbits) on $\CP^1$.
We note that a generic $\Gamma$-orbit has $|\Gamma|/2$ points in it,
and they form a one-complex-parameter family.
This means that the dimension of the space of degree-$|\Gamma|/2$ semi-invariants is two.
This also means that any homogeneous semi-invariant is a product of a number of degree-$|\Gamma|/2$ semi-invariants and of semi-invariants associated from special orbits.

The special orbits of binary tetrahedral/octahedral/icosahedral groups $\Gamma_{T/O/I}$ on $\CP^1\simeq S^2$
correspond to the vertices, the middle point of the edges, and the barycenters of the faces of a tetrahedron, octahedron or icosahedron, respectively.
They give rise to the following semi-invariants:
\begin{equation}
\begin{array}{c|ccc}
&\text{vertex} & \text{edge} & \text{face}\\
\hline
\Gamma_T & \varphi_4 & \varphi_6 & \varphi_4' \\
\Gamma_O& \varphi_6 & \varphi_{12} & \varphi_8\\
\Gamma_I & \varphi_{12} & \varphi_{30} & \varphi_{20}
\end{array}.
\end{equation}
Since generic orbits consist of $|\Gamma_T|/2=12$, $|\Gamma_O|/2=24$, $|\Gamma_I|/2=60$ points, we see that they can be normalized so that \begin{align}
T:&&(\varphi_4)^3+(\varphi_4')^3+(\varphi_6)^2&=0,\\
O:&&(\varphi_6)^4+(\varphi_8)^3+(\varphi_{12})^2&=0,\\
I:&&(\varphi_{12})^5+(\varphi_{20})^3+(\varphi_{30})^2&=0.
\end{align}
The three generators and the one relation describe the ring of semi-invariants.\if0\footnote{%
By the way, I studied implications of these semi-invariants in the AdS$_5$/CFT$_4$ correspondence long time ago \cite{Tachikawa:2007tt}. 
As nobody cites it, let me cite it here. I do not think it was a particularly bad paper.
}\fi

Let us now come back to the questions of the invariants of $G$.
Since elements of $G$ and elements of $\Gamma$ are different only up to an overall $U(1)$ phase,
invariants of $G$ are necessarily semi-invariants of $\Gamma$.
We know the invariants of $G$ form a free polynomial ring,
and we know the ring of semi-invariants of $\Gamma$ explicitly.
A moment of thought reveals that in each case, the generating invariants of $G$ can be taken to be a suitable powers of suitable generators of generators of semi-invariants of $\Gamma$.
This explains the patterns of degrees of invariants of rank-2 complex reflection groups shown in Table~\ref{table}.

Let us also mention that these semi-invariants of $\Gamma$ allow us to determine the invariant ring of $\Gamma$ itself.
We simply quote the results from \cite{Springer}, in the form $\bC[V]^\Gamma=\bC[X,Y,Z]/\text{eq}$, where
\begin{equation}
\begin{array}{c|ccc|r@{\,}l}
& X&Y&Z&\multicolumn{2}{c}{\text{eq}} \\
\hline
T& 2^{-1/2} \varphi_6 &  -\varphi_4 \varphi_4' & i(\varphi_4^3 +\frac12\varphi_6^2) & X^4+Y^3 + Z^2&=0 \\
O& \varphi_8 & \varphi_6^2 & \varphi_{6}\varphi_{12} & X^3Y+Y^3+Z^2&=0\\
I& \varphi_{12} & \varphi_{20} & \varphi_{30} & X^5+Y^3+Z^2&=0
\end{array}
\end{equation}
The icosahedral case is particularly simple in that the semi-invariants are in fact invariants.
This follows from the fact that the binary icosahedral group is perfect and therefore has no nontrivial one-dimensional representation.

\bibliographystyle{ytphys}
\baselineskip=.95\baselineskip
\bibliography{ref}

\end{document}